\documentclass[twocolumn,showpacs,preprintnumbers,amsmath,amssymb]{revtex4}
\def\eps{\epsilon}

\def\3nab{\tilde{\nabla}}

\def\hs {\,-\,}

\def\be {\begin{equation}}
\def\ee {\end{equation}}
\def\bea {\begin{eqnarray}}
\def\eea {\end{eqnarray}}

\usepackage{graphicx}% Include figure files
\usepackage{dcolumn}% Align table columns on decimal point
\usepackage{bm}% bold math
\usepackage{longtable}
\begin{document}
%%%%%%%%%%%%%%%%%%%%%%%%%%%%%%%%%%%%%%%%%%%%%%%%%%%%
\title{Exponential Potentials on the Brane}
%%%%%%%%%%%%%%%%%%%%%%%%%%%%%%%%%%%%%%%%%%%%%%%%%%%%
\author{Naureen Goheer$^1$ and  Peter K.\ S.\ Dunsby$^{1,2}$}
\affiliation{1. Department of Mathematics and Applied Mathematics,
  University of Cape Town, 7701 Rondebosch, Cape Town, South Africa}
\affiliation{2. South African Astronomical Observatory, 
  Observatory 7925, Cape Town, South Africa.}
%%%%%%%%%%%%%%%%%%%%%%%%%%%%%%%%%%%%%%%%%%%%%%%%%%%%% 
\begin{abstract}
In this paper we study the cosmological dynamics of Randall\hs Sundrum 
braneworld type scenarios in which the five\hs dimensional Weyl tensor has 
a non\hs vanishing projection onto the three\hs brane where matter fields are 
confined.  Using dynamical systems techniques, we study how the state 
space of Friedmann-Lema\^{\i}tre-Robertson-Walker (FLRW) and Bianchi 
type I scalar field models with an exponential potential is affected by the 
bulk Weyl tensor, focusing on the differences that appear with respect to 
standard general relativity and also Randall\hs Sundrum cosmological scenarios 
without the Weyl tensor contribution.
\end{abstract}

\maketitle
%%%%%%%%%%%%%%%%%%%%%%%%%%%%%%%%%%%%%%%%%%%%%%%%%%%%%%
\section{Introduction}
%%%%%%%%%%%%%%%%%%%%%%%%%%%%%%%%%%%%%%%%%%%%%%%%%%%%%%
The notion that we live on a three\hs dimensional brane embedded in a
higher\hs dimensional spacetime has attracted a considerable amount 
of interest over the last few years. These ideas have had a long history 
(for references see~\cite{Visser}) but its recent revival is due to recent 
work by Randall and Sundrum~\cite{Randall}.

In such braneworld scenarios the ordinary matter is confined to the brane
while gravity can propagate in the whole spacetime (known as the bulk). The 
effective four\hs dimensional gravity on the brane is modified by extra
terms~\cite{Binetruy,Shiromizu} in the gravitational equations, 
one quadratic in the energy\hs momentum tensor and the other representing the 
electric part of the five\hs dimensional Weyl tensor.

The dynamics of a braneworlds filled with a perfect fluid have been
intensively investigated during last three years. In particular 
Campus and Sopuerta \cite{Campos1,Campos2} have recently studied the complete 
dynamics of Friedmann\hs Lema\^{\i}tre\hs Robertson\hs Walker (FLRW) 
and the Bianchi type I and V cosmological models with a barotropic 
equation of state  using an approach first introduced by by Goliath and 
Ellis \cite{Goliath} (see also \cite{Dynamics} for a detailed 
discussion of the application of dynamical systems techniques to 
cosmology).

Their analysis led to the discovery of new critical points corresponding 
to the Bin$\acute{\mbox{e}}$truy\hs Deffayet\hs Langlois (BDL) models 
\cite{Binetruy}, representing the evolution at very high energies, 
where effects due to the extra dimension become dominant. These solutions
appear to be a generic feature of the state space of more general 
cosmological models. They also showed that the state space contains new 
bifurcations, demonstrating how the dynamical character of some of the 
critical points changes relative to the general\hs relativistic case.
Finally, they showed that for models satisfying all the ordinary energy 
conditions, causality requirements and for $gamma>1$, the anisotropy is 
negligible near the initial singularity, a result first demonstrated 
by Maartens {\it et. al.} \cite{Maartens-Sahni}.

Scalar field dynamics on the brane is considerably more interesting due to 
the fact that the equation of state parameter $\gamma$ is now dynamical. In 
a recent paper \cite{Paper1} we considered the dynamics of inflationary models 
with an exponential potential $V(\phi)=\exp(b\phi)$, an important class
of inflationary models first considered by Burd and Barrow \cite{Burd} 
and Haliwell \cite{Halliwell} in the context of standard general relativity 
(GR). These models have a richer dynamical structure in the brane\hs world 
scenario and have the nice feature that inflation can take place with potentials 
ordinarily too steep to sustain inflation which result from the high\hs energy 
corrections to the Friedmann equation \cite{Copeland2}. Indeed for $b<0$ we found 
past attractors representing high energy {\it steep inflationary} models in which 
inflation ends naturally as the energy drops below the brane tension and the condition 
for inflation no longer holds.

In this paper we complete this analysis (hereafter refereed to as Paper I), 
studying the effects of the five\hs dimensional Weyl tensor on FLRW and 
Bianchi type I cosmological models. In the case of FLRW models this study 
will be completely general whereas in the Bianchi type I case we will 
neglect (or constrain) the Weyl tensor components for which the theory does 
not provide evolution equations. The paper is organized as follows. 
In section~\ref{sec:prelims} we briefly outline the geometric formulation 
of brane\hs world scenarios and give the main dynamical equations for FLRW and 
Bianchi type I cosmological models where the matter is described by a dynamical 
scalar field $\phi$ with an exponential potential $V(\phi)=\exp(b\phi)$.  
The dynamics of the FLRW and Bianchi type I cosmological models are presented in 
section~\ref{sec:FLRW} and~\ref{sec:Bianchi} respectively. Finally we end with a discussion 
of the main results of the analysis in section~\ref{sec:Conclusion}. Through this paper 
we follow the following notation: upper\hs case Latin indices denote coordinates in the 
bulk spacetime ($A,B,\ldots = 0 - 4$) whereas lower\hs case Latin indices 
denote coordinates on the brane ($a,b,\ldots=0 - 3$).  We use physical units in 
which $c=1$.
%%%%%%%%%%%%%%%%%%%%%%%%%%%%%%%%%%%%%%%%%%%%%%%%%%%%
\section{Preliminaries} \label{sec:prelims}
%%%%%%%%%%%%%%%%%%%%%%%%%%%%%%%%%%%%%%%%%%%%%%%%%%%%
To begin with we introduce the geometrical framework of braneworld
cosmological models and the main assumptions used to study FLRW and 
Bianchi type I cosmological models.
%%%%%%%%%%%%%%%%%%%%%%%%%%%%%%%%%%%%%%%%%%%%%%%%%%%%%%%%%%%%%%%%%%
\subsection{Basic equations of the brane\hs world}
%%%%%%%%%%%%%%%%%%%%%%%%%%%%%%%%%%%%%%%%%%%%%%%%%%%%%%%%%%%%%%%%%%
In Randall\hs Sundrum brane\hs world type scenarios matter fields are
confined in a three\hs brane embedded in a five\hs dimensional 
spacetime (bulk). It is assumed that the metric of this spacetime,
$g^{(5)}_{AB}$, obeys the Einstein equations with a negative 
cosmological constant $\Lambda_{(5)}$ \cite{Shiromizu,Sasaki,Maartens})
\begin{equation}
G^{(5)}_{AB} = -\Lambda_{(5)}g^{(5)}_{AB}
+\kappa^2_{(5)} \delta(\chi)\left[ -\lambda\,g_{AB}+T_{AB}
\right]\;,  \label{Einstein}
\end{equation}
where $G^{(5)}_{AB}$ is the Einstein tensor, $\kappa_{(5)}$ denotes 
the five\hs dimensional gravitational coupling constant and  
$T_{AB}$ represents the energy\hs momentum tensor of the matter with the 
Dirac delta function reflecting the fact that matter is confined 
to the spacelike hypersurface $x^4\equiv\chi=0$ (the brane) with induced 
metric $g_{AB}$ and tension $\lambda$.   

Using the Gauss\hs Codacci equations, the Israel junction conditions and
the $Z_2$ symmetry with respect to the brane the effective  
Einstein equations on the brane are
\begin{equation}
G_{ab}=-\Lambda g_{ab}+\kappa^2 T_{ab}+
\kappa^4_{(5)} S_{ab} - {\cal E}_{ab} \,, \label{Modified}
\end{equation}
where $G_{ab}$ is the Einstein tensor of the induced metric $g_{ab}$. 
The four\hs dimensional gravitational constant $\kappa$ and the 
cosmological constant $\Lambda$ can be expressed in terms of the 
fundamental constants in the bulk $(\kappa_{(5)},\Lambda_{(5)})$ and 
the brane tension $\lambda $ \footnote{In order to recover conventional 
gravity on the brane $\lambda$ must be assumed to be positive.}.

As mentioned in the introduction, there are two corrections to the 
general\hs relativistic equations. Firstly $S_{ab}$ represent corrections 
quadratic in the matter variables due to the form of the Gauss\hs Codacci 
equations:
\begin{equation}
S_{ab} = \textstyle{\frac{1}{12}}T T_{ab}-\textstyle{\frac{1}{4}}
T_a{}^c T_{bc}+\textstyle{\frac{1}{24}}g_{ab}\left[3T^{cd}
T_{cd}-T^2\right] \;. \label{Sab}
\end{equation}
Secondly ${\cal E}_{ab}$, corresponds to the ``electric'' part of the 
five\hs dimensional Weyl tensor $C^{(5)}_{ABCD}$ with respect to the 
normals, $n_A$ ($n^An_A=1$), to the hypersurface $\chi=0$, that is
\begin{equation}
{\cal E}_{AB} = C^{(5)}_{ACBD}n^Cn^D \;,
\end{equation}
representing the non\hs local effects from the free gravitational 
field in the bulk.
The modified Einstein equations (\ref{Modified}) together with the 
conservation of energy\hs momentum equations $\nabla^a T_{ab}=0$ lead
to a constraint on $S_{ab}$ and ${\cal E}_{ab}$:
\begin{equation} 
\nabla^a ({\cal E}_{ab}-\kappa^4_{(5)}S_{ab}) = 0 \,. \label{Conserved}
\end{equation}
We can decompose ${\cal E}_{ab}$ into its ineducable parts relative 
to any timelike observers with 4\hs velocity $u^a$ ($u^au_a=-1$):
\begin{equation} 
{\cal E}_{ab} = -\left(\frac{\kappa^{}_{(5)}}{\kappa}\right)^4
\left[ (u_au_b+\textstyle{\frac{1}{3}}h_{ab}){\cal U} +2u_{(a}{\cal Q}_{b)}
+{\cal P}_{ab}\right]\,, \label{spli}
\end{equation}
where 
\begin{equation}
{\cal Q}_au^a=0\,,~~~ {\cal P}_{(ab)}= {\cal P}_{ab}\,,~~
{\cal P}^a{}_a=0\,,~~ {\cal P}_{ab}u^b=0 \,. 
\end{equation}
Here ${\cal U}$ has the same form as the energy\hs momentum tensor 
of a radiation perfect fluid and for this reason is referred to
as the ``dark'' energy density of the Weyl fluid. ${\cal Q}_a$ is a 
spatial and ${\cal P}_{ab}$ is a spatial, symmetric and trace\hs 
free tensor.  ${\cal Q}_a$ and ${\cal P}_{ab}$ are analogous to the usual 
energy flux vector $q^a$ and anisotropic stress tensor $\pi_{ab}$ in 
General Relativity. The constraint equation (\ref{Conserved}) leads to 
evolution equations for ${\cal U}$ and ${\cal Q}_a$, but not 
for ${\cal P}_{ab}$ (see \cite{Maartens}).
%%%%%%%%%%%%%%%%%%%%%%%%%%%%%%%%%%%%%%%%%%%%%%%%%%%%
\subsection{Scalar field dynamics on the brane}
%%%%%%%%%%%%%%%%%%%%%%%%%%%%%%%%%%%%%%%%%%%%%%%%%%%%
In this paper we consider both  FLRW and Bianchi I cosmological models on 
the brane where the matter is described by a dynamical scalar field $\phi$ with 
an exponential potential $V(\phi)=\exp(b\phi)$. In this case the fluid 4\hs 
velocity can be written as 
\begin{equation}
u^a=-\frac{\nabla_a\phi}{\dot{\phi}}\;,~~u^au_a=-1\;,
\label{Four-velocity}
\end{equation}
which makes it automatically orthogonal to the hypersurfaces of homogeneity
(surfaces of constant $\phi$). With this choice of 4\hs velocity, the 
energy\hs momentum tensor $T_{\mu\nu}$ for a scalar field $\phi$ 
takes the form of a perfect fluid (See page 17 in \cite{BED} for details):
\begin{equation}
T_{\mu\nu}=\rho u_{\mu}u_{\nu}+ph_{\mu\nu}\,,\label{Scalar-EMT}
\end{equation}
with
\begin{equation}
\rho=\frac{1}{2}\dot{\phi}^2+V(\phi)\label{Scalar-energy}
\end{equation}
and 
\begin{equation}
p=\frac{1}{2}\dot{\phi}^2-V(\phi)\;.\label{Scalar-momentum}
\end{equation}
where $\dot{\phi}$ is the momentum density of the scalar field and 
$V(\phi)$ is its potential energy. If the scalar field is not 
minimally coupled this simple representation is no longer valid, 
but it is still possible to have an imperfect fluid form for the 
energy\hs momentum tensor \cite{Madsen}.

In the case of FLRW models, the effective Einstein Equations (\ref{Modified})
lead to the conditions
\[ {\cal Q}_a = {\cal P}_{ab} = 0 \,, \]
and this, through the constraint~(\ref{Conserved}), further implies
\begin{equation}  
D_a{\cal U}=0 ~ \Leftrightarrow ~ {\cal U}={\cal U}(t)\,,\label{grau}  
\end{equation}
where $D_a$ denotes the covariant derivative associated with the
induced metric on the hypersurfaces of homogeneity ($h_{ab}\equiv
g_{ab}+u_au_b$).  The situation is somewhat different when one 
considers Bianchi type I models. This time we obtain ${\cal Q}_a=0$ but 
we do not get any restriction on ${\cal P}_{ab}\,.$  Since there is no way 
of fixing the dynamics of this tensor we will study the particular case in 
which it vanishes and the conditions~(\ref{grau}) also apply.

The key equations describing the dynamics of these models are the modified
Raychaudhuri equation 
\begin{eqnarray}
\dot{H} & = & -H^2-\frac{3\gamma-2}{6}\kappa^2\rho\left[1+\frac{3\gamma-1}
{3\gamma-2}\frac{\rho}{\lambda}\right]-\frac{2}{3}\sigma^2 \nonumber \\
& & +\frac{1}{3}\Lambda-\frac{2{\cal U}}{\lambda\kappa^2}
\,,\label{ray}
\end{eqnarray}
which is an evolution equation for the Hubble parameter 
$H {\textstyle\frac{1}{3}}\equiv \nabla_a u^a$, and the 
Friedmann equation constraint (a first integral of (\ref{ray}): 
\begin{equation} 
H^2 = \frac{1}{3}\kappa^2\rho\left(1+\frac{\rho}{2\lambda}\right)
-\frac{1}{6}{}^3R+\frac{1}{3}\sigma^2+\frac{1}{3}\Lambda +
\frac{2{\cal U}}{\lambda\kappa^2}\,,
\label{Friedmann}
\end{equation}
where ${}^3R$ is the scalar curvature of the hypersurfaces orthogonal
to the fluid flow and $2\sigma^2\equiv\sigma^{ab}\sigma_{ab}$ is the
shear scalar ($\sigma_{ab}\equiv h_a{}^ch_b{}^d\nabla_{(c}u_{d)}-
Hh_{ab}$).  We consider only the case of a positive four\hs dimensional
cosmological constant, i.e. $\Lambda\geq 0$.   
For Bianchi type I models ${}^3R$ vanishes whereas for FLRW
models it is given by ${}^3R=6ka^{-2}(t)\,.$ On the other hand,
the shear vanishes for FLRW models and for Bianchi type I models 
the evolution of the shear scalar is
\[ (\sigma^2)^\cdot = -6H\sigma^2 \,. \]
In both cases the evolution equation for ${\cal U}$ is given by~\cite{Maartens}
\[ \dot{\cal U} = - 4H{\cal U}\,, \]
and substituting for $\rho$ and $p$ from (\ref{Scalar-energy}) and 
(\ref{Scalar-momentum}) into the energy conservation equation 
\begin{equation}
\dot{\rho}+\Theta(\rho +p)=0\;,
\label{Energy-conservation} 
\end{equation}
leads to the 1+3 form of the Klein\hs Gordon equation 
\begin{equation}
\ddot{\phi} +\Theta \dot{\phi} +V'(\phi)=0\;, \label{Klein-Gordon}
\end{equation}
an exact ordinary differential equation for $\phi$ once the potential
has been specified.
It is convenient to relate $p$ and $\rho$ by the {\em index $\gamma$} 
defined by 
\begin{equation}
p = ( \gamma - 1) \rho  ~~\Leftrightarrow ~~ \gamma = \frac{p + \rho}{\rho}
~=~\frac{\dot{\phi}^2}{\rho}\;.
\end{equation}
This index would be constant in the case of a simple one\hs component 
fluid, but in general will vary with time in the case of a scalar 
field:
\begin{equation}
{\dot{\gamma}}=\Theta\gamma(\gamma-2)-2\gamma\frac{V'}{\dot{\phi}}\;.
\label{Gammadot}
\end{equation}
Notice that this equation is well\hs defined even for $\dot{\phi}\rightarrow 0$,
since $\frac{\gamma}{\dot{\phi}}=\frac{\dot{\phi}}{\rho}$.

In the next two sections we study in detail the dynamics of 
(a) FLRW models with ${\cal U} \neq 0$ and (b) the anisotropic 
Bianchi I models with ${\cal U} \neq 0$ thus extending our recent work  
which only considered FLRW models with vanishing non\hs local energy 
density ${\cal U}$. The key difference between this work and a 
similar study carried out by Campos and Sopuerta \cite{Campos2}
is that we have a dynamical equation of state parameter $\gamma$ 
which evolves according to equation (\ref{Gammadot}) above.
In the case of an exponential potential $V(\phi)=\exp(b\phi)$, 
where $b\leq 0$ we obtain
\begin{equation}
\dot{\gamma}=3H\gamma(\gamma-2)+\sqrt{3\gamma}(\gamma-2)b
\sqrt{\frac{\rho^2}{3\kappa^2}}\;.
\label{gammadot}
\end{equation}
In order to obtain a compact state space we will re\hs write the above 
dynamical equations in terms of dimensionless coordinates that are 
appropriately expansion normalized.   
%%%%%%%%%%%%%%%%%%%%%%%%%%%%%%%%%%%%%%%%%%%%%%%%%%%%%%%%%%%%%%%
\section{Analysis of FLRW models with exponential potentials} \label{sec:FLRW}
%%%%%%%%%%%%%%%%%%%%%%%%%%%%%%%%%%%%%%%%%%%%%%%%%%%%%%%%%%%%%%%
In the first paper of this series \cite{Paper1} we had to distinguish 
between the case ${}^3R\leq 0$ and ${}^3R\geq 0$ when introducing appropriate 
expansion normalized coordinates for the FLRW models. In this paper, we have 
to consider four different subcases when studying the FLRW models and each one
will have to be normalized by different quantities in order to obtain a compact 
state space:  
(A) ${\cal U}\geq 0$ and ${}^3R\leq 0$;  
(B) ${\cal U}\geq 0$ and ${}^3R\geq 0$; (C) ${\cal U}\leq 0$ and 
${}^3R\leq 0$; (D) ${\cal U}\leq 0$ and ${}^3R\geq 0$.
The total state space is composed of these 4 sectors, which are disconnected 
since trajectories cannot leave one sector and enter another. For a detailed 
description of the state space, in particular the invariant sets and the 
unusual geometry of sector D we refer to \cite{Campos2}. 
%%%%%%%%%%%%%%%%%%%%%%%%%%%%%%%%%%%%%%%%%%%%%%%%%%%%%%%%%%%%%%%
\subsection{${\cal U} \geq 0 $ and $^3R \leq 0$} \label{sec:FLRWA}
%%%%%%%%%%%%%%%%%%%%%%%%%%%%%%%%%%%%%%%%%%%%%%%%%%%%%%%%%%%%%%%
As explained by several authors (see e.g. \cite{Goliath,Campos2}), a compact 
state space is obtained by using the dimensionless general\hs relativistic 
variables
\begin{equation}
\Omega_\rho\equiv \frac{\kappa^2\rho}{3H^2}\,,   ~~ 
\Omega_k\equiv -\frac{{}^3R}{6H^2}\,,   ~~ 
\Omega_\Lambda\equiv \frac{\Lambda}{3H^2} \label{def1} 
\end{equation}
together with the following new variables:
\begin{equation}
\Omega_\lambda\equiv \frac{\kappa^2\rho^2}{6\lambda H^2} 
\,,  ~~ \Omega_{\cal U} \equiv  \frac{2{\cal U}}{\lambda\kappa^2H^2} 
\,. \label{def2}
\end{equation}
The coordinates represent the fractional contributions of the ordinary energy 
density, the curvature, the brane tension, and the dark radiation energy to the 
total energy density, normalized with respect to the Hubble parameter $H(t)$. 
In addition, we use the variable $\gamma\in[0,2]$ in order to describe the 
dynamics of the scalar field $\phi$.  Using these variables, the 
Friedmann constraint (\ref{Friedmann}) becomes:
\be
\label{frieda}
\Omega_{\rho}+\Omega_{k}+\Omega_\Lambda+\Omega_{\lambda}+\Omega_{\cal{U}}=1\,.
\ee
Since all of the terms in the sum are non\hs negative, each of the variables 
$\Omega_\rho,~\Omega_k,~\Omega_\lambda,~\Omega_{\cal U}$ takes values 
in the interval $[0,1]$, which means that we obtain a compact state space. 

In order to decouple the evolution equation for $H$ (\ref{ray}) 
from the evolution equations for the density parameters 
$\Omega_{i}$, we introduce the dimensionless time derivative
\begin{equation} 
'\equiv |H|^{-1}d/dt\,.
\end{equation} 
Inserting the Friedmann constraint (\ref{frieda}) in order to keep 
the dimensionality of the state space as low as possible, it follows 
that the dynamics of the open or flat models with positive 
$\cal{U}$ are described by the following equations:
\bea 
\gamma '&=&\eps \sqrt{3\gamma} (\gamma -2)[\sqrt{3\gamma}
+\eps~ b \sqrt{1-\Omega_{k}-\Omega_\Lambda-
\Omega_{\lambda}-\Omega_{\cal{U}}}]\;,\nonumber \\
\Omega_{k}'&=&\eps[(3\gamma-2)(1-\Omega_{k})
+3\gamma(\Omega_{\lambda}-\Omega_\Lambda)+(4-3\gamma)\Omega_{\cal{U}}]
\Omega_{k}\;, \nonumber \\
\Omega_\Lambda'&=&\eps[3\gamma(1+\Omega_{\lambda}-\Omega_\Lambda-\Omega_{k}
-\Omega_{\cal{U}})
+2\Omega_{k}+4\Omega_{\cal{U}}]\Omega_\Lambda\;,  
\label{systema} \\
\Omega_{\lambda}'&=&\eps[3\gamma(\Omega_{\lambda}-\Omega_\Lambda-\Omega_{k}
-1-\Omega_{\cal{U}})
+2\Omega_{k}+4\Omega_{\cal{U}}]\Omega_{\lambda}\;, \nonumber \\
\Omega_{\cal{U}}'&=&\eps[(3\gamma-4)(1-\Omega_{\cal{U}})+(2-3\gamma)\Omega_{k}+
3\gamma(\Omega_{\lambda}-\Omega_\Lambda)]\Omega_{\cal{U}}\;.
\nonumber
\eea
The equilibrium points of this dynamical system, their coordinates in state 
space and their eigenvalues are given in TABLE I below.
%%%%%%%%%%%%%%%%%%%%%%%%%%%%%%%%%%%%%%%%%%%%%%
%TABLE; EIGENVALUES A
%%%%%%%%%%%%%%%%%%%%%%%%%%%%%%%%%%%%%%%%%%%%%%
\begin{table*}
\caption{\label{tab:table1} This table gives the coordinates and 
eigenvalues of the critical points with ${\cal U} \geq 0 $ and 
$^3R \leq 0$. 
We have defined $\psi=\sqrt{\frac{8}{b^2}-3},~\zeta=\sqrt{\frac{64}{b^2}-15}$.}
\begin{ruledtabular}
\begin{tabular}{ccc}
Model  & Coordinates   & Eigenvalues \\ \hline
$\mbox{F}_\epsilon^2$ & $(2,0,0,0,0)$ & 
$(6\epsilon+\sqrt{6}b,4\eps,6\eps,-6\eps,2\eps)$ \\
$\mbox{M}_\epsilon^0$ & $(0,1,0,0,0)$ & non-hyperbolic \\
$\mbox{M}_\epsilon^2$ & $(2,1,0,0,0)$ & $2\epsilon(3,-2,1,-5,-1)$ \\
$\mbox{dS}_\epsilon^2$ & $(2,0,1,0,0)$ & 
$(6\epsilon,-2\eps,-6\eps,-12\eps,-4\eps)$ \\
$\mbox{m}_\epsilon^0(\Omega_{\Lambda},\Omega_{\lambda})~for~ b=0$ \footnote
{The eigenvalues of these points can only be evaluated for $\Omega_{\Lambda}
+\Omega_{\lambda}\neq 1$. For $\Omega_{\Lambda}+\Omega_{\lambda}=1$ a 
perturbative analysis has to be carried out.} & 
$(0,0,\Omega_{\Lambda},\Omega_{\lambda},0)$ & $-2\epsilon(3,1,0,0,2)$ \\
$\mbox{m}_\epsilon^0(\Omega_{\Lambda},\Omega_{\lambda})~for~ b 
\neq 0~^a$ & $(0,0,\Omega_{\Lambda},\Omega_{\lambda},0)$ & 
$(\infty\footnote{This actually reads 
$lim_{\gamma \rightarrow 0}(-6\eps-\sqrt{3}b\sqrt{\frac{1-\Omega_{\lambda}-
\Omega_{\Lambda}}
{\gamma}})=\infty$.},
-2\epsilon,0,0,-4\eps)$ \\
$\mbox{m}_\epsilon^2$ & $(2,0,0,1,0)$ & $2\epsilon(3,5,6,3,4)$ \\
$\mbox{R}_\epsilon^0$ & $(0,0,0,0,1)$ & non-hyperbolic \\
$\mbox{R}_\epsilon^2$ & $(2,0,0,0,1)$ & $2\epsilon(3,1,2,-4,-1)$ \\
$\mbox{F}_{+}^{b^2/3}~~\footnote{$b^2\in[0,6]$}$ & $(\frac{b^2}{3},0,0,0,0)$ & 
$(\frac{b^2}{2}-3,b^2-2,b^2,-b^2,b^2-4)$\\
$\mbox{X}_{+}^{2/3}(b)~~\footnote{$b^2\in[2,\infty[$}$ & $(\frac{2}{3},
1-\frac{2}{b^2},0,0,0)$ & $(-1-\psi,-1+\psi,2,-2,-2)$\\
$\mbox{A}_{+}^{4/3}(b)~~\footnote{$b^2\in[4,\infty[$}$ & $(\frac{4}{3},0,0,0,
1-\frac{4}{b^2})$ & $(-\frac{1}{2}(1+\zeta),2,4,-4,-\frac{1}{2}(1-\zeta))$\\
\end{tabular}
\end{ruledtabular}
\end{table*}
Most of the equilibrium points of the system (\ref{systema}) are the
same as those obtained in Paper I \cite{Paper1}: the flat FLRW universe 
$F_{\epsilon}^2$ with stiff matter and $a(t)=t^{1/3}$; the Milne universe 
$M_{\epsilon}^2$ with stiff matter and  $a(t)=t$; the Milne universe 
with $\gamma=0$ ($M_{\epsilon}^0$); the flat non\hs general\hs relativistic 
model $m_{\epsilon}^{2}$ with $\gamma=2$ and scale factor $a(t)=t^{1/6}$, 
which has been discussed in detail in \cite{Campos1}; the flat FLRW universe 
$F_{+}^{b^2/3}$ with $\gamma=\frac{b^2}{3}$ and $a(t)=t^{2/b^2}$; 
and a set of universe models $X_{+}^{2/3}(b)$ with $\gamma=\frac{2}{3}$ and curvature 
$\Omega_{k}=1-\frac{2}{b^{2}}$ depending on the value of b.

Furthermore we find the de\hs Sitter model $dS_\eps^2$ with $\gamma=2$ and 
scale factor $a(t)=exp(\sqrt{\Lambda/3}t)$, and a set of 
non\hs general\hs relativistic critical points with $\gamma=0$ which extends 
both in $\Omega_{\Lambda}$\hs  and $\Omega_{\lambda}$\hs direction. This set is 
denoted by $m_{\epsilon}^0(\Omega_{\Lambda},\Omega_{\lambda})$; it contains 
the flat general relativistic FLRW model and the general relativistic 
de\hs Sitter model with constant energy density $\rho$ respectively. 
This 2\hs parameter set of equilibrium points contains the line 
of non\hs general\hs relativistic critical points $m_{\epsilon}^0(\Omega_{\lambda})$ 
obtained in our previous paper \cite{Paper1}. 

In addition to these equilibrium points with vanishing $\cal{U}$, we find the 
following points with ${\cal U }> 0$: the points $R_\epsilon^0$ and $R_\epsilon^2$ which
have the same metric $a(t)=t^{1/2}$ as the flat the FLRW model with $\gamma=\frac{4}{3}$ 
(radiation), but have $\gamma=0,~2$ respectively and $\rho=k=\lambda^{-1}=0$. 
The point ${A}_{+}^{4/3}(b)$ describes a flat expanding model with 
$\gamma=\frac{4}{3}$ and vanishing brane\hs tension ($\Omega_{\lambda}=0$), 
but which has in general non\hs vanishing energy density and non\hs local 
energy density contributions $\Omega_\rho,~\Omega_{\cal{U}}>0$. 
For $b=0$, this model has maximal positive $\cal U$ ($\Omega_{\cal{U}}=1,~\Omega_{\rho}=0$), 
whereas for $b^2=4$ the model coincides with the flat FLRW model with 
$\gamma=4/3$ ($\Omega_\rho=1,~\Omega_{\cal{U}}=0$). The scale factor of this 
model is proportional to $t^{1/2}$ for all values of $b$.

Notice that $F_{+}^{b^2/3}$ only occurs for $0\leq b^{2}\leq 6$, 
$X_{+}^{2/3}(b)$ occurs in this sector of state space for $b^{2}\geq 2$, and 
${A}_{+}^{4/3}(b)$ occurs only for $b^2\geq 4$ in this sector of state space. 
All three points  move in state space as we vary the parameter value $b$, but  
independent of the value of $b$  only occur in the expanding sector $\epsilon=+1$.

Note that equilibrium points $M_{\epsilon}^0,~m_{\epsilon}^0(\Omega_\Lambda,
\Omega_{\lambda})$ and $R_\eps^0$ are non\hs hyperbolic; we analyzed their 
nature using the perturbative methods described in Paper 1 \cite{Paper1}.
%%%%%%%%%%%%%%%%%%%%%%%%%%%%%%%%%%%%%%%%%%%%%%%%
\subsection{${\cal U} \geq 0$ and $^3R \geq 0$}
%%%%%%%%%%%%%%%%%%%%%%%%%%%%%%%%%%%%%%%%%%%%%%%%
Here we will use the variables $\gamma,Q,\tilde{\Omega}_{\rho},\tilde{\Omega}
_\Lambda,\tilde{\Omega}_{\lambda},\tilde{\Omega}_{\cal{U}}$, where 
we define 
\bea
Q \equiv \frac{H}{D},~ with~D^2 \equiv H^2+{\textstyle\frac{1}{6}}{^3R}
\eea 
and the variables with a tilde are the analogues of those in (\ref{def1}) 
and (\ref{def2}) but normalized with respect to $D$ instead of $H$.  
The Friedmann equation (\ref{Friedmann}) then becomes 
\be
\label{friedb}
\tilde{\Omega}_{\rho}+\tilde{\Omega}_\Lambda+\tilde{\Omega}_{\lambda}+
\tilde{\Omega}_{\cal{U}}=1\,,
\ee
from which it can be seen that the quantities $\tilde{\Omega}_i$ take 
values in $[0,1]$, whereas $Q$ takes values in $[-1,1]$. Hence these 
coordinates define a compact state space. Introducing the time derivative
\begin{equation}
\nonumber
' \equiv D^{-1}d/{dt}\;,
\end{equation}
the evolution equation for $D$ decouples from the evolution equations for 
$\tilde{\Omega}_i$. Inserting the Friedmann constraint (\ref{friedb}), we 
find that the dynamics of the closed and flat models are described by     
\bea
\gamma '&=&\sqrt{3\gamma} (\gamma -2)[\sqrt{3\gamma}Q + b 
\sqrt{1-\tilde{\Omega}_{\lambda}-\tilde{\Omega}_\Lambda-\tilde{\Omega}_
{\cal{U}}}]\;,
\nonumber\\
Q'&=&[1-\frac{3}{2}\gamma(1+\tilde{\Omega}_{\lambda}-\tilde{\Omega}_\Lambda
-\tilde{\Omega}_{\cal{U}})
-2\tilde{\Omega}_{\cal{U}}](1-Q^2)\;,
\nonumber\\
\tilde{\Omega}_{\Lambda}'&=&[3\gamma 
(1+\tilde{\Omega}_{\lambda}-\tilde{\Omega}_\Lambda-\tilde{\Omega}_{\cal{U}})+
4\tilde{\Omega}_{\cal{U}}]Q \tilde{\Omega}_{\Lambda}\;,
\nonumber\\
\tilde{\Omega}_{\lambda}'&=&[3\gamma 
(\tilde{\Omega}_{\lambda}-\tilde{\Omega}_\Lambda-1-\tilde{\Omega}_{\cal{U}})+
4\tilde{\Omega}_{\cal{U}}]Q \tilde{\Omega}_{\lambda}\;,
\nonumber\\
\tilde{\Omega}_{\cal{U}}'&=&[(3 \gamma -4)(1-\tilde{\Omega}_{\cal{U}})+
3\gamma (\tilde{\Omega}_{\lambda}-\tilde{\Omega}_\Lambda)]
Q \tilde{\Omega}_{\cal{U}}\;.
\label{systemb}
\eea
The equilibrium points of this system, their coordinates in state space and 
their eigenvalues are given in TABLE II. 
%%%%%%%%%%%%%%%%%%%%%%%%%%%%%%%%%%%%%%%%%%%%%%%%%%%%%
%TABLE; EIGENVALUES B
%%%%%%%%%%%%%%%%%%%%%%%%%%%%%%%%%%%%%%%%%%%%%%%%%%%%%
\begin{table*}
\caption{\label{tab:table2} This table gives the coordinates and 
eigenvalues of the critical points with ${\cal U} \geq 0$ and
 $^3R \geq 0$. 
We have defined the real quantities $\chi=\sqrt{\frac{8-3b^2}{2}},\xi=
\sqrt{2}b(\frac{3}{2}-\frac{1}{\gamma})\sqrt{3\gamma(1-\tilde{\Omega}_\Lambda-
\tilde{\Omega}_{\cal U})+
2\tilde{\Omega}_{\cal U}-1},
~\alpha=\sqrt{9\gamma^2(\tilde{\Omega}_\Lambda+\tilde{\Omega}_{\cal U}-1)+6
\gamma(1-2\tilde{\Omega}_{\cal U})+4(1-\tilde{\Omega}_{\cal U})}$. 
Notice that $\alpha$ is a real positive 
quantity within the allowed parameter range.}
\begin{ruledtabular}
\begin{tabular}{ccc}
Model  & Coordinates   & Eigenvalues \\ \hline
$\mbox{F}_\epsilon^2$ & $(2,\eps,0,0,0)$ &
 $(6\epsilon+\sqrt{6}b,4\eps,6\eps,-6\eps,2\eps)$ \\
$\mbox{dS}_\epsilon^2$ & $(2,\eps,1,0,0)$ &
 $(6\eps,-2\eps,-6\eps,-12\eps,-4\eps)$ \\
$\mbox{m}_\epsilon^0(\tilde{\Omega}_{\Lambda},\tilde{\Omega}_{\lambda})
~for~ b=0~^a$ & 
$(0,\eps,\tilde{\Omega}_{\Lambda},\tilde{\Omega}_{\lambda},0)$ 
& $-2\eps(3,1,0,0,2)$ \\
$\mbox{m}_\epsilon^0(\tilde{\Omega}_{\Lambda},\tilde{\Omega}_{\lambda})~
for~ b \neq 0$ \footnote
{The eigenvalues of these points can only be evaluated for 
$\tilde{\Omega}_{\Lambda}+\tilde{\Omega}_{\lambda}\neq 1$. For 
$\tilde{\Omega}_{\Lambda}+\tilde{\Omega}_{\lambda}= 1$ a 
perturbative analysis has to be carried out.} & 
$(0,\eps,\tilde{\Omega}_{\Lambda},\tilde{\Omega}_{\lambda},0)$
 & $(\infty\footnote{This actually reads 
$lim_{\gamma \rightarrow 0}(-6\eps-\sqrt{3}b\sqrt{\frac{1-\tilde{\Omega}_
{\lambda}-\tilde{\Omega}_
{\Lambda}}{\gamma}})=\infty$.},-2\epsilon,0,0,-4\eps)$ \\
$\mbox{R}_\epsilon^0$ & $(0,\eps,0,0,1)$ & non-hyperbolic  \\
$\mbox{m}_\epsilon^2$ & $(2,\eps,0,1,0)$ & $2\epsilon(3,5,6,3,4)$ \\
$\mbox{R}_\epsilon^2$ & $(2,\eps,0,0,1)$ & $2\epsilon(3,1,2,-4,-1)$ \\
$\mbox{E}$ & $(\gamma^\ast,0,\tilde{\Omega}^\ast_{\Lambda},
\tilde{\Omega}^\ast_\lambda,\tilde{\Omega}^\ast_{\cal U})$ 
& $(\xi,\alpha,0,0,-\alpha)$ \\

$\mbox{F}_{+}^{b^2/3}~~\footnote{$b^2\in[0,6]$} $ & $(\frac{b^2}{3},1,0,0,0)$ 
& $(\frac{b^2}{2}-3,b^2-2,b^2,-b^2,b^2-4)$\\
$\mbox{X}_{+}^{2/3}(b)~~\footnote{$b^2\in[0,2]$}$ & 
$(\frac{2}{3},-\frac{b}{\sqrt{2}},0,0,0)$ 
 & $(\frac{b}{\sqrt{2}}-\chi,\frac{b}{\sqrt{2}}
+\chi,-\sqrt{2}b,\sqrt{2}b,\sqrt{2}b)$ \\
$\mbox{A}_{+}^{4/3}(b)~~\footnote{$b^2\in[4,\infty[$}$ & 
$(\frac{4}{3},1,0,0,1-\frac{4}{b^2})$ 
& $(-\frac{1}{2}(1+\zeta),2,4,-4,-\frac{1}{2}(1-\zeta))$ \\
\end{tabular}
\end{ruledtabular}
\end{table*}
We recover the equilibrium points obtained in the previous subsection which 
corresponded to flat models. In particular, for $b^2\geq 4$ we find the 
equilibrium point $A_+^{4/3}(b)$, which for $b^2\neq 4$ describes a model with 
non\hs vanishing non\hs local energy density $\cal{U}$. 

In addition, we find a 2\hs parameter set of static ($Q=0$) models $E$. These 
models occur for any fixed values of coordinates $(\gamma^\ast,0,
\tilde{\Omega}^\ast_\Lambda,
\tilde{\Omega}^\ast_\lambda,\tilde{\Omega}^\ast_{\cal U})$ 
subject to the constraints (\ref{friedb}) and 
\begin{equation} 
1-\frac{3}{2}\gamma^\ast(1+\tilde{\Omega}^\ast_\lambda-\tilde{\Omega}^\ast
_\Lambda-\tilde{\Omega}^\ast_{\cal U})
-2\tilde{\Omega}^\ast_{\cal U}=0  \label{Econ1}
\end{equation}
as well as 
\begin{equation} 
\sqrt{\gamma^\ast}(\gamma^\ast-2)b\sqrt{1-\tilde{\Omega}^\ast_\Lambda-
\tilde{\Omega}^\ast_\lambda-
\tilde{\Omega}^\ast_{\cal U}}=0 \,. 
\label{Econ2}
\end{equation}
Again, we have already inserted the Friedmann constraint (\ref{friedb}). Note 
that the Jacobian of the dynamical system 
(\ref{systembs}) is in general not well\hs defined for $\gamma=0$ 
\footnote{For $\gamma=b=0$ 
the Jacobian is well\hs defined and the non\hs zero eigenvalues of $nE$ are given by 
$\pm \alpha\mid_{\gamma=0}$.}. In that case a perturbative study confirms that 
the points $E\mid_{\gamma^\ast=0}$ are indeed saddle points in state space.

The models $E$ have the same line\hs element as the Einstein universe 
($H=0,~k=+1$), but in general also non\hs zero brane tension and positive 
nonlocal energy density $\cal U$. The quantities $a,\rho,\cal U$ are 
constants $a^\ast,\rho^\ast,{\cal U}^\ast$.

Requiring positive energy density $\rho^\ast$, we find from (\ref{Friedmann}) 
that the constants $a^\ast,{\cal U}^\ast$ must satisfy 
\begin{equation}
{\cal U}^\ast \leq \frac{\lambda \kappa^2}{2} (\frac{1}{a^{\ast^2}}
-\frac{\Lambda}{3}) \,.
\end{equation}

Notice that the constraint (\ref{Econ2}) comes from the $\gamma'$\hs 
equation in (\ref{systemb}), hence from the dynamics of the scalar field, and 
does not occur in the studies of a similar model with constant equation of 
state, i.e. without a dynamical equation of state parameter. 

From (\ref{Econ1}) we can see that the static models $E$ can occur for 
any values of $\gamma\in[0,2]$. Equation (\ref{Econ2}) is however in 
general only satisfied for all values of $\gamma$ for the vacuum models 
($\tilde{\Omega}_{\rho}=0$) \footnote{If the potential is flat ($b=0$) 
equation(\ref{Econ2}) is satisfied for any values of $\gamma,~\tilde{\Omega}_{\rho}$. 
We will not discuss this case in much detail since it is physically not very interesting.}.

Note that in contrast to dynamical models, it is not possible to have 
$\tilde{\Omega}_{\rho}=0$ and $\tilde{\Omega}_{\lambda}\neq0$ when considering
 static models. This is due to the fact that for static models $a=a^\ast,
~\rho=\rho^\ast$ are constants and therefore $\tilde{\Omega}_{\rho}=0$ implies
 $\rho^\ast=0$, which leads to $\tilde{\Omega}_{\lambda}=0$. This has the 
interesting consequence that there exist no vacuum static solutions in the 
$\Lambda=0$\hs subset. Only when allowing for the additional degree of freedom
 $\tilde{\Omega}_{\Lambda}$ do we find that (\ref{Econ1}) can be satisfied 
for $\tilde{\Omega}_{\rho}=0$.

We emphasize that the static model denoted by $E^{1/3}$ in our previous paper 
\cite{Paper1} is in that sense unphysical, since that model found in the 
$\Lambda=0$ case has $Q=0,~\tilde{\Omega}_{\rho}=0$ and $\tilde{\Omega}_{\lambda}=1$ 
which can only be satisfied in this extended scenario allowing for non\hs zero 
cosmological constant $\Lambda$.

Hence the only static vacuum model occurring in this sector is the model with 
$\tilde{\Omega}_{\lambda}=0$, which means that $\tilde{\Omega}_{\cal U}=1/2,
~\tilde{\Omega}_{\Lambda}=1/2$. This is a line of equilibrium points extending 
in $\gamma$\hs direction. The other static models are the ones with $\gamma=0$ or
  $\gamma=2$. The former are characterized by $\tilde{\Omega}_{\cal U}=1/2$, 
the latter have $\tilde{\Omega}_{\cal U}=2+3(\tilde{\Omega}_\lambda-
\tilde{\Omega}_{\Lambda})$.  

Note that the only static models with $\tilde{\Omega}_\Lambda=0$ are the ones 
with $\gamma=0,~\tilde{\Omega}_{\cal U}=1/2,~\tilde{\Omega}_\lambda+
\tilde{\Omega}_\rho=1/2$.

For constant potential ($b=0$) equation (\ref{Econ2}) is automatically fulfilled 
for all values of $\gamma^\ast,\tilde{\Omega}^\ast_{\Lambda},
\tilde{\Omega}^\ast_\lambda,\tilde{\Omega}^\ast_{\cal U}$. Thus for $b=0$ $E$ 
degenerates into a 3\hs dimensional surface containing static models for all 
values of $\gamma\in[0,2]$. 

Notice that in the $\tilde{\Omega}_{\Lambda}=0$\hs subset much stronger 
constraints must again be satisfied; it can be shown that here $\gamma\in
[0,\frac{2}{3}]$ is required even for $b=0$.

Although these sets of equilibrium points form geometrically interesting 
objects in state space, they are not of interest for our stability analysis, 
since all these equilibrium points are unstable saddle points. 
%%%%%%%%%%%%%%%%%%%%%%%%%%%%%%%%%%%%%%%%%%%%%%%%
\subsection{${\cal U} \leq 0$ and $^3R \leq 0$} \label{sec:FLRWC}
%%%%%%%%%%%%%%%%%%%%%%%%%%%%%%%%%%%%%%%%%%%%%%%%
In this case we obtain a compact state space by introducing the dynamical 
variables $Z,\bar{\Omega}_\rho,\bar{\Omega}_k,\bar{\Omega}_{\Lambda},
\bar{\Omega}_\lambda$, where
\begin{equation} 
Z\equiv\frac{H}{N}\,,~~N^2 \equiv H^2
-\frac{2{\cal U}}{\lambda\kappa^2}\,,\label{defz}
\end{equation}
and the variables $\bar{\Omega}_i$ are defined as in  in~(\ref{def1}) 
and~(\ref{def2})
but normalized with respect to $N$ instead of $H$.
Using these variables, the Friedmann constraint reads
\be
\label{friedc}
\bar{\Omega}_{\rho}+\bar{\Omega}_{k}+\bar{\Omega}_{\Lambda}+
\bar{\Omega}_{\lambda}=1
\ee  
As before, all the terms in that sum are non\hs negative by definition, hence 
the variables $\bar{\Omega}_\rho,\bar{\Omega}_k,\bar{\Omega}_{\Lambda},
\bar{\Omega}_\lambda$ take values in the interval $[0,1]$. 
Since $\gamma\in[0,2]$ and $Z\in[-1,1]$, we find that the state 
space is again compact.

Furthermore, introducing the time derivative 
\begin{equation}
'\equiv N^{-1}d/dt\,,
\end{equation}
the evolution of $N$ decouples from the rest of the variables. We obtain
\bea
\gamma '&=& \sqrt{3\gamma} (\gamma -2)[\sqrt{3\gamma} Z + 
b \sqrt{1-\bar{\Omega}_{\lambda}-\bar{\Omega}_{k}-\bar{\Omega}_{\Lambda}}]\;,
\nonumber\\
Z'&=&-[\frac{3}{2} \gamma (1+\bar{\Omega}_{\lambda}-\bar{\Omega}_{\Lambda})+(
1-\frac{3}{2} 
\gamma )\bar{\Omega}_{k}-2](1-Z^2)\;,
\nonumber\\
\bar{\Omega}_{k}'&=&[(3\gamma -2)(1-\bar{\Omega}_{k})+
3\gamma (\bar{\Omega}_{\lambda}-\bar{\Omega}_{\Lambda})]Z \bar{\Omega}_{k}\;,
\nonumber\\
\bar{\Omega}_{\Lambda}'&=&[3\gamma(1-\bar{\Omega}_{k}-\bar{\Omega}_{\Lambda}+
\bar{\Omega}_{\lambda})+2 \bar{\Omega}_k]Z \bar{\Omega}_{\Lambda}\;,
\nonumber\\
\bar{\Omega}_{\lambda}'&=&[2\bar{\Omega}_{k}-
3\gamma (1+\bar{\Omega}_{k}+\bar{\Omega}_{\Lambda}-\bar{\Omega}_{\lambda})] 
Z \bar{\Omega}_{\lambda}\,.
\label{systemc}
\eea
The equilibrium points of this system, their coordinates in state space and 
their eigenvalues are given in TABLE III.
%%%%%%%%%%%%%%%%%%%%%%%%%%%%%%%%%%%%%%%%%%%%%%%%%%%%%
%TABLE; EIGENVALUES C
%%%%%%%%%%%%%%%%%%%%%%%%%%%%%%%%%%%%%%%%%%%%%%%%%%%%%
\begin{table*}
\caption{\label{tab:table3} This table gives the coordinates and 
eigenvalues of the critical points with ${\cal U} \leq 0$ and 
$^3R \leq 0$. 
We have defined $\psi=\sqrt{\frac{8}{b^2}-3},~\xi=\sqrt{64-15b^2},\eta=
\sqrt{2}b(\frac{3}{2}-\frac{1}{\gamma})\sqrt{3\gamma(1-\bar{\Omega}_\Lambda-
\bar{\Omega}_k)+
\bar{\Omega}_{k}-2},
~\Gamma=\sqrt{9\gamma^2(\bar{\Omega}_\Lambda+\bar{\Omega}_k-1)+6
\gamma(2-\bar{\Omega}_k)+4(1-\bar{\Omega}_k)}$. Notice that $\Gamma$ is a real 
positive quantity within the allowed range of variables.}
\begin{ruledtabular}
\begin{tabular}{ccc}
Model  & Coordinates   & Eigenvalues \\ \hline
$\mbox{F}_\epsilon^2$ & $(2,\eps,0,0,0)$ & $(6\epsilon+\sqrt{6}b,2\eps,4\eps,
6\eps,-6\eps)$ \\
$\mbox{M}_\epsilon^0$ & $(0,\eps,1,0,0)$ & non-hyperbolic \\
$\mbox{M}_\epsilon^2$ & $(2,\eps,1,0,0)$ & $2\epsilon(3,-1,-2,1,-5)$ \\
$\mbox{dS}_\epsilon^2$ & $(2,\eps,0,1,0)$ & $(6\epsilon,-4\eps,-2\eps,
-6\eps,-12\eps)$ \\
$\mbox{m}_\epsilon^0(\bar{\Omega}_{\Lambda},\bar{\Omega}_{\lambda})
~for~ b=0$ \footnote
{The eigenvalues of these points can only be evaluated for 
$\bar{\Omega}_{\Lambda}+\bar{\Omega}_{\lambda}\neq 1$. For 
$\bar{\Omega}_{\Lambda}+\bar{\Omega}_{\lambda}= 1$ a 
perturbative analysis has to be carried out.} & 
$(0,\eps,0,\bar{\Omega}_{\Lambda},\bar{\Omega}_{\lambda})$ & 
$-2\epsilon(3,2,1,0,0)$ \\
$\mbox{m}_\epsilon^0(\bar{\Omega}_{\Lambda},\bar{\Omega}_{\lambda})
~for~ b 
\neq 0~^a$ & $(0,\eps,0,\bar{\Omega}_{\Lambda},\bar{\Omega}_{\lambda})$ & 
$(\infty\footnote{This 
actually reads $lim_{\gamma \rightarrow 0}(-6\eps-\sqrt{3}b\sqrt
{\frac{1-\bar{\Omega}_{\Lambda}-\bar{\Omega}_{\lambda}}{\gamma}})=\infty$.},
-4\eps,-2\epsilon,0,0)$ \\
$\mbox{m}_\epsilon^2$ & $(2,\eps,0,0,1)$ & $2\epsilon(3,4,5,6,3)$ \\
$\mbox{S}$ & 
$(\gamma^\ast,0,\bar{\Omega}_k^\ast,\bar{\Omega}^\ast_{\Lambda},\bar{\Omega}_
{\lambda}^\ast)$ & $(\eta,\Gamma,0,-\Gamma,0)$ \\
$\mbox{F}_{+}^{b^2/3}$ & $(\frac{b^2}{3},1,0,0,0)$ & 
$(\frac{b^2}{2}-3,b^2-4,b^2-2,b^2,-b^2)$\\
$\mbox{X}_{+}^{2/3}(b)~~\footnote{$b^2 \in [2,\infty[$} $ & $(\frac{2}{3},1,
1-\frac{2}{b^2},0,0)$ 
& $(-1-\psi,-2,-1+\psi,2,-2)$\\
$\mbox{A}_{+}^{4/3}(b)~~\footnote{$b^2 \in [0,4]$} $ & $(\frac{4}{3},-\frac{b}
{2},0,0,0)$ 
& $(\frac{1}{4}(b-\xi),\frac{1}{4}(b+\xi),-b,-2b,2b)$\\
\end{tabular}
\end{ruledtabular}
\end{table*}
Again we recover the models with ${\cal{U}}=0$ that we have obtained in the 
previous two sections. Notice that the point $A_+^{4/3}(b)$ now only occurs 
for $b^2\in[0,4]$. This is due to the fact that the point represents a model 
with  nonlocal energy density depending on the value of the parameter $b$. 
For $b^2<4$, the point corresponds to a model with negative nonlocal 
energy density $\cal{U}$, whereas for $b^2>4$ it represents a model with 
positive $\cal{U}$. For $b^2=4$, the point coincides with the expanding FLRW 
model ${F}_{+}^{b^2/3}\mid_{b^2=4}$ with ${\cal U}=0$ and 
$\gamma=\frac{4}{3}$. Thus the point $A_+^{4/3}(b)$ moving 
in state space is leaving the sector describing models with ${\cal{U}}\leq 0$ 
and entering the sector of models with ${\cal{U}} \geq 0$ when the parameter 
value $b^2=4 ~(b=-2)$.

In addition to these points, we find the set $S$ of static models $(H=0)$ 
whose coordinates 
$(\gamma^\ast,0,\bar{\Omega}^\ast_k,\bar{\Omega}^\ast_{\Lambda},
\bar{\Omega}^\ast_{\lambda})$ satisfy the 
constraints
\begin{equation} 
2-\frac{3}{2}\gamma^\ast(1+\bar{\Omega}^\ast_\lambda-
\bar{\Omega}^\ast_k-\bar{\Omega}^\ast_{\Lambda})-\bar{\Omega}^\ast_k=0 \, 
\label{Scon1}
\end{equation}
and 
\begin{equation} 
\sqrt{\gamma^\ast}(\gamma^\ast-2)b\sqrt{1-\bar{\Omega}^\ast_k-
\bar{\Omega}^\ast_{\Lambda}-\bar{\Omega}^\ast_\lambda}=0 \,, 
\label{Scon2}
\end{equation}
where we have already inserted the Friedmann constraint (\ref{friedc}).

Requiring non\hs negative energy density $\rho^\ast$, we also find from 
(\ref{Friedmann}) that the constants
$a^\ast,{\cal U}^\ast$ must satisfy 
\begin{equation} 
-{\cal U}^\ast \geq \frac{\lambda\kappa^2}{2}(\frac{\Lambda}{3}-
\frac{k}{a^{\ast^2}}) \,. 
\label{}
\end{equation}
It is important to note that these static models in general have negative 
curvature ($k=-1$).

We can see from (\ref{Scon1}) and (\ref{friedc}) that there are no static 
vacuum models. The only static models for $b\neq0$ are the ones with 
$\gamma^\ast=2$ and $\bar{\Omega}_k^\ast=\frac{1}{2}+\frac{3}{2}(\bar{\Omega}_
{\lambda}^\ast-\bar{\Omega}^\ast_{\Lambda})$. These models occur both in the 
$\bar{\Omega}_{\Lambda}=0$-subset and outside of that region. Again this is 
a 2\hs parameter surface. 

For $b=0$, this 2\hs dimensional set degenerates into a 3\hs dimensional surface; 
in this special case we find that static models occur for all values 
of $\gamma^\ast \geq 2/3$.

Notice that the constraints on these models, in contrast to the constraints on 
the models $E$ found in the previous section, do not change when we restrict 
ourselves to the $\Lambda=0$\hs subset: for $b=0$ there are static models for all 
$\gamma\geq 2/3$ even for $\bar{\Omega}_{\Lambda}=0$. 

Dynamically these models are again not very interesting, since they all 
represent saddle points in state space.
%%%%%%%%%%%%%%%%%%%%%%%%%%%%%%%%%%%%%%%%%%%%%%%%
\subsection{${\cal U} \leq 0$ and $^3R \geq 0$}
%%%%%%%%%%%%%%%%%%%%%%%%%%%%%%%%%%%%%%%%%%%%%%%%
As explained in \cite{Campos2}, we need to take into account that the 
Friedmann equation~(\ref{Friedmann}) now has two non\hs positive terms.  
It turns out that we can obtain a compact state space introducing the 
dimensionless dynamical variables $W,\hat{\Omega}_\rho,\hat{\Omega}_{\Lambda},
\hat{\Omega}_\lambda,
\hat{\Omega}_{\cal U}$, where
\[ W\equiv \frac{H}{P}\,,~~~P^2 \equiv H^2+
\frac{1}{6}{}^3R-\frac{2{\cal U}}{\lambda\kappa^2} \,,\]
and the variables with a hat are defined as those in~(\ref{def1}) and 
(\ref{def2}) but normalized with respect to $P$ instead of $H$.

The Friedmann constraint now reduces to
\be
\label{friedd}
\hat{\Omega}_{\rho}+\hat{\Omega}_{\Lambda}+\hat{\Omega}_{\lambda}=1\;.
\ee  
It is important to note that $\hat{\Omega}_{\cal U}$ does not 
appear in the Friedmann equation (\ref{friedd}). Indeed, it can be seen 
from its definition that $\hat{\Omega}_{\cal U}$ is negative and belongs to 
the interval $[-1,0]$. As before, we can see from (\ref{friedd}) that 
$\hat{\Omega}_{\rho},\hat{\Omega}_{\Lambda},\hat{\Omega}_{\lambda}\in[0,1]$. 
Together with $W\in[-1,1]$ and $\gamma\in [0,2]$, we find that these variables 
define a compact state space.

Using the time derivative 
\begin{equation}
' \equiv P^{-1}d/dt\,,
\end{equation}
we obtain
\bea
\gamma '&=& \sqrt{3\gamma} (\gamma -2)[\sqrt{3\gamma}W + 
b \sqrt{1-\hat{\Omega}_{\lambda}-\hat{\Omega}_{\Lambda}}]\;,
\nonumber\\
W'&=&[1-\frac{3}{2}\gamma(1+\hat{\Omega}_{\lambda}-\hat{\Omega}_{\Lambda})]
(1-W^2)-\hat{\Omega}_{\cal{U}}\;,
\nonumber\\
\hat{\Omega}_{\Lambda}'&=&3\gamma 
(1+\hat{\Omega}_{\lambda}-\hat{\Omega}_{\Lambda})W \hat{\Omega}_{\Lambda}\;,
\nonumber\\
\hat{\Omega}_{\lambda}'&=&3\gamma 
(\hat{\Omega}_{\lambda}-\hat{\Omega}_{\Lambda}-1)W \hat{\Omega}_{\lambda}\;,
\nonumber\\
\hat{\Omega}_{\cal{U}}'&=&[3\gamma (1+\hat{\Omega}_{\lambda}-\hat{\Omega}_
{\Lambda})-4] W \hat{\Omega}_{\cal{U}}\,.
\label{systemd}
\end{eqnarray}
The equilibrium points of this system, their coordinates in state space and 
their eigenvalues are given in TABLE IV.
%%%%%%%%%%%%%%%%%%%%%%%%%%%%%%%%%%%%%%%%%%%%%%%%%%%%%
%TABLE; EIGENVALUES D
%%%%%%%%%%%%%%%%%%%%%%%%%%%%%%%%%%%%%%%%%%%%%%%%%%%%%
\begin{table*}
\caption{\label{tab:table4} This table gives the coordinates and 
eigenvalues of the critical points with ${\cal U} \leq 0$ and 
$^3R \geq 0$. 
We have defined $\chi=\sqrt{\frac{8-3b^2}{2}}$ and $\xi=\sqrt{64-15b^2},
~\delta=\sqrt{6}b\sqrt{1-\hat{\Omega}_{\Lambda}-\hat
{\Omega}_{\lambda}},~\beta=\sqrt{9\gamma^2{\hat{\Omega}_
{\lambda}}+4}$. Notice that $\beta$ is a 
real positive quantity within the allowed range of variables.}
\begin{ruledtabular}
\begin{tabular}{ccc}
Model  & Coordinates   & Eigenvalues \\ \hline
$\mbox{F}_\epsilon^2$ & $(2,\eps,0,0,0)$ & 
$(6\epsilon+\sqrt{6}b,4\eps,6\eps,-6\eps,2\eps)$ \\
$\mbox{dS}_\epsilon^2$ & $(2,\eps,1,0,0)$ & 
$(6\epsilon,-2\eps,-6\eps,-12\eps,-4\eps)$ \\
$\mbox{m}_\epsilon^0(\hat{\Omega}_{\Lambda},\hat{\Omega}_{\lambda})~
for~ b=0$ \footnote
{The eigenvalues of these points can only be evaluated for 
$\hat{\Omega}_{\Lambda}+\hat{\Omega}_{\lambda}\neq 1$. For 
$\hat{\Omega}_{\Lambda}+\hat{\Omega}_{\lambda}= 1$ a 
perturbative analysis has to be carried out.}& 
$(0,\eps,\hat{\Omega}_{\Lambda},\hat{\Omega}_{\lambda},0)$ 
& $-2\epsilon(3,1,0,0,2)$ \\
$\mbox{m}_\epsilon^0(\hat{\Omega}_{\Lambda},\hat{\Omega}_{\lambda})
~for~ b 
\neq 0~^a$ & $(0,\eps,\hat{\Omega}_{\Lambda},\hat{\Omega}_{\lambda},0)$ & 
$(\infty\footnote{This 
actually reads $lim_{\gamma \rightarrow 0}(-6\eps-\sqrt{3}b\sqrt{\frac{1-\hat
{\Omega}_{\Lambda}-\hat
{\Omega}_{\lambda}}{\gamma}})=\infty$.},-2\epsilon,0,0,-4\eps)$ \\

$\mbox{m}_\epsilon^2$ & $(2,\eps,0,1,0)$ & $2\epsilon(3,5,6,3,4)$ \\
$\mbox{E}$ & $(\gamma^\ast,0,\hat{\Omega}^\ast_{\Lambda},\hat{\Omega}^\ast
_{\lambda},\hat{\Omega}^\ast_
{\cal U})$ & $(\delta,\beta,0,-\beta,0)$ \\
$\mbox{F}_{+}^{b^2/3}$ & $(\frac{b^2}{3},1,0,0,0)$ & 
$(\frac{b^2}{2}-3,b^2-2,b^2,-b^2,b^2-4)$\\
$\mbox{X}_{+}^{2/3}(b)~~\footnote{$b^2 \in [0,2]$}$ & $(\frac{2}{3},-\frac{b}
{\sqrt{2}},0,0,0)$ 
& $(\frac{b}{\sqrt{2}}-\chi,\frac{b}{\sqrt{2}}+\chi,-\sqrt{2}b,
\sqrt{2}b,\sqrt{2}b)$\\
$\mbox{A}_{+}^{4/3}(b)~~\footnote{$b^2 \in [0,4]$}$ & $(\frac{4}{3},-\frac{b}
{2},0,0,\frac{b^2}{4}-1)$ & 
$(\frac{1}{4}(b-\xi),\frac{1}{4}(b+\xi),-b,-2b,2b)$\\
\end{tabular}
\end{ruledtabular}
\end{table*}
Again, we recover the equilibrium points that represent flat models obtained 
in the previous subsection. In addition, we find another set of equilibrium 
points representing static models denoted by $E$.

These models correspond to the Einstein\hs universe\hs like models found in 
subsection B, since they also have the same line element as the Einstein 
universe ($H=0,~k=+1$) and have non\hs zero brane tension and non\hs local 
energy density $\cal U$, except now ${\cal U}\leq 0$.

Their coordinates $(\gamma^\ast,0,\hat{\Omega}^\ast_\Lambda,
\hat{\Omega}^\ast_\lambda,
\hat{\Omega}^\ast_{\cal U})$ satisfy the constraints 
\begin{equation}
\label{econ1}
1-\frac{3}{2}\gamma^\ast(1+\hat{\Omega}^\ast_\lambda-\hat{\Omega}^\ast_
\Lambda)-\hat{\Omega}^\ast_{\cal U}=0
\end{equation}
and 
\begin{equation}
\label{econ2}
\sqrt{\gamma^\ast}(\gamma^\ast-2)b\sqrt{1-\hat{\Omega}^\ast_\Lambda-
\hat{\Omega}^\ast_\lambda}=0\,.
\end{equation}
where we have again inserted the Friedmann constraint (\ref{friedd}).

We can see from (\ref{econ1}) that there exist no vacuum solutions and no 
solutions for $\gamma^\ast=0$. Hence for a non\hs flat potential the only static 
models are the ones with $\gamma^\ast=2$ and $\hat{\Omega}^\ast_{\cal U}=
-2-3(\hat{\Omega}^\ast_\lambda-\hat{\Omega}^\ast_\Lambda)$. In particular there
 are no static models for $\hat{\Omega}^\ast_\Lambda=0$.

For $b=0$, we find static models for all $\gamma\geq 1/3$; the ones in the 
$\hat{\Omega}^\ast_\Lambda=0$\hs subset occur for $\gamma\in[1/3,4/3]$. 
%%%%%%%%%%%%%%%%%%%%%%%%%%%%%%%%%%%%%
\subsection{Qualitative Analysis}
%%%%%%%%%%%%%%%%%%%%%%%%%%%%%%%%%%%%%
We use the results obtained in the previous subsections in order to determine 
the dynamical character of the equilibrium points found above. We summarize 
the results in TABLE V below.   
%%%%%%%%%%%%%%%%%%%%%%%%%%%%%%%%%%%%%%%%%%%%%%
%TABLE; DYNAMICAL CHARACTER D
%%%%%%%%%%%%%%%%%%%%%%%%%%%%%%%%%%%%%%%%%%%%%%
\begin{table*}
\caption{\label{tab:table5} Dynamical character of the critical
points in the FLRW case for $\Lambda=0$.}
\begin{ruledtabular}
\begin{tabular}{ccccccccccc}
Model  & $b=0$ & $0<b^2<2$ & $b^2=2$ & $2<b^2<4$ & $b^2=4$ & $4<b^2<6$ & 
$b^2=6$ & $b^2>6$ \\ \hline
$\mbox{m}_{+}^0(\Omega_{\lambda})$ & sink & saddle & saddle
& saddle & saddle & saddle & saddle & saddle \\ 
$\mbox{m}_{-}^0(\Omega_{\lambda})$ & source & source &
 source &  source &  source & source & source & source\\
$\mbox{E}$ & saddle & saddle & saddle & saddle & saddle & saddle & saddle & 
saddle \\
$\mbox{S}$ & saddle & saddle & saddle & saddle & saddle & saddle & saddle & 
saddle \\ 
$\mbox{F}_{\eps}^2$ & saddle & saddle & saddle & 
saddle & saddle & saddle & saddle & saddle\\
$\mbox{m}_{+}^2$ & source & source & source & 
source & source & source & source & source \\
$\mbox{m}_{-}^2$ & sink & sink & sink & sink & sink & sink & sink & sink\\
$\mbox{R}_\epsilon^0$ & saddle & saddle & saddle & 
saddle & saddle & saddle & saddle & saddle \\
$\mbox{R}_\epsilon^2$ & saddle & saddle & saddle & 
saddle & saddle & saddle & saddle & saddle \\
$\mbox{F}_{+}^{b^2/3}$ & sink & sink & saddle 
\footnote{Notice that this point is an attractor for all general relativistic 
open or flat 
models, while it is a repeller for all closed general relativistic models.}  
& saddle & 
saddle & saddle & saddle & -\\
$\mbox{X}_{+}^{2/3}(b)$ & saddle & saddle & saddle $^a$ & sink & sink & sink 
& sink & 
sink\\
$\mbox{A}_{+}^{4/3}(b)$\footnote{Notice that this model has negative non-local 
energy density $\cal{U}$ for $b^2<4$, ${\cal{U}}=0$ for $b^2=4$ and 
${\cal{U}}>0$ for $b^2>4$} & saddle & saddle & saddle $^a$ & saddle & saddle 
& saddle & saddle & saddle\\
\end{tabular}
\end{ruledtabular}
\end{table*}
%%%%%%%%%%%%%%%%%%%%%%%%%%%%%%%%%%%%%%%%%%%%%%
%TABLE; DYNAMICAL CHARACTER D
%%%%%%%%%%%%%%%%%%%%%%%%%%%%%%%%%%%%%%%%%%%%%%
\begin{table*}
\caption{\label{tab:table6} Dynamical character of the critical
points in the FLRW case for general $\Lambda$.}
\begin{ruledtabular}
\begin{tabular}{ccccccccccc}
Model  & $b=0$ & $0<b^2<2$ & $b^2=2$ & $2<b^2<4$ & $b^2=4$ & $4<b^2<6$ & 
$b^2=6$ & $b^2>6$ \\ \hline
$\mbox{m}_{+}^0(\Omega_{\Lambda},\Omega_{\lambda})$ & sink & saddle &
saddle & saddle & saddle & saddle & saddle & saddle \\ 
$\mbox{m}_{-}^0(\Omega_{\Lambda},\Omega_{\lambda})$ & source & source &
 source & source & source & source & source & source\\
$\mbox{E}$ & saddle & saddle & saddle & saddle & saddle & saddle & saddle & 
saddle \\
$\mbox{S}$ & saddle & saddle & saddle & saddle & saddle & saddle & saddle & 
saddle \\ 
$\mbox{F}_{\eps}^2$ & saddle & saddle & saddle & 
saddle & saddle & saddle & saddle & saddle\\
$\mbox{dS}_\epsilon^2$ & saddle & saddle & saddle & 
saddle & saddle & saddle & saddle & saddle \\
$\mbox{m}_{+}^2$ & source & source & source & 
source & source & source & source & source \\
$\mbox{m}_{-}^2$ & sink & sink & sink & sink & sink & sink & sink & sink\\
$\mbox{R}_\epsilon^0$ & saddle & saddle & saddle & 
saddle & saddle & saddle & saddle & saddle \\
$\mbox{R}_\epsilon^2$ & saddle & saddle & saddle & 
saddle & saddle & saddle & saddle & saddle \\
$\mbox{F}_{+}^{b^2/3}$ & sink & saddle & saddle & saddle & 
saddle & saddle & saddle & -\\
$\mbox{X}_{+}^{2/3}(b)$ & saddle & saddle & saddle & saddle & saddle & 
saddle & saddle & saddle\\
$\mbox{A}_{+}^{4/3}(b)$\footnote{Notice that this model has negative non-local 
energy density $\cal{U}$ for $b^2<4$, ${\cal{U}}=0$ for $b^2=4$ and 
${\cal{U}}>0$ for $b^2>4$} & saddle & saddle & saddle & saddle & saddle 
& saddle & saddle & saddle\\
\end{tabular}
\end{ruledtabular}
\end{table*}
For all values of $b$ the past attractors of the FLRW models are the BDL 
models $m_+^2$ with $\gamma=2~(V<<\dot{\phi}^2)$ 
and the set of non\hs general relativistic models $m_+^0(\Omega_\Lambda,
\Omega_\lambda)$ with $\gamma=0~ (\dot{\phi}^2<<V)$. 
For a non\hs flat potential ($b\neq0$) the contracting BDL model $m_-^2$ with 
$\gamma=2$ is the unique future attractor. In 
particular the expanding deSitter model is {\it not} a future attractor in the 
presence of a scalar field with exponential potential. This stands in strong 
contrast to the results obtained in \cite{Campos2} for the scalar field\hs free case 
where the expanding de\hs Sitter model $dS_+$ is a future attractor for all 
values of $\gamma$. In the special case of a 
flat potential $(b=0)$ we find that the models $m_+^0(\Omega_\Lambda,
\Omega_\lambda)$ including the de\hs Sitter model with 
$\gamma=0$ form another set of future attractors. Note that in the 
$\Lambda=0$-subset the general-relativistic models 
$F_+^{b^2/3}$ and $X_+^{2/3}(b)$ are also future attractors. To be precise,
the flat FLRW model $F_+^{b^2/3}$ is an 
attractor for $b^2<2$ and the model $X_+^{2/3}(b)$ for $b^2>2$.

Notice that in General Relativity, the scenario that we are describing here 
(matter described by a dynamical scalar field with exponential potential) only 
admits a static universe in the special case of a constant potential 
($b=0$). This model had the same line element as the Einstein universe 
($H=0,~k=+1)$ and only occurred at $\gamma=2/3$. 

If we allow for non\hs zero brane tension but neglect the bulk effects 
(${\cal U} =0$, see Paper I (\cite{Paper1})), this condition is relaxed: 
we find static universe models characterized by $H=0, ~k=+1$ for all values 
of $\gamma\in]1/3,2/3]$ for $b=0$. For a non\hs flat potential $(b\neq0)$ there 
are still no physical static equilibrium points \footnote{As explained above, 
the static model $E^{1/3}$ with $\tilde{\Omega}_\rho=0$ and $\tilde{\Omega}_\lambda=1$ 
must be excluded.}.
 
In this paper taking into account the bulk effects as well as allowing for a 
non\hs zero cosmological constant $\Lambda$, we find that there are 
not only static Einstein universe like models ($k=+1$), but also static 
saddle points that are flat ($k=0$) or even negatively curved ($k=-1$). 

For $b=0$ there are static models for all values of $\gamma$ even in the 
$\Lambda=0$-subset, which shows that these are purely due to the bulk effects. 
Allowing for $\Lambda\neq0$ we find that the constraints on the 
static models are further relaxed. We then find that Einstein static models 
occur for all values of $\gamma$ when ${\cal U}\geq0$ and for all 
$\gamma\geq 1/3$ when ${\cal U}\leq0$. Flat and open static models occur for all
 $\gamma\geq 2/3$ but only if ${\cal U}\leq 0$.

More interestingly we also find static models for $b\neq0$. For $\Lambda=0$ we 
find Einstein static models with $\gamma=0$ in the ${\cal U}\geq0$-sector and 
for $\gamma=2$ there are open and flat models in the ${\cal U}\leq0$-sector. 

When allowing for a cosmological constant, we find that the Einstein static 
universe occurs for all values of $\gamma$ when ${\cal U}\geq0$ and for 
$\gamma=2$ when ${\cal U}\leq0$. The open and flat models still only occur for 
$\gamma=2$ and only in the ${\cal U}\leq0$\hs sector.

Altogether, the conditions for allowing for static models in General 
Relativity are changed dramatically when considering the terms corresponding to 
brane tension and non\hs local energy density, and are further relaxed when 
including a cosmological constant. Instead of finding only the static Einstein 
universe with $\gamma=\frac{2}{3}$, and this only in the special case of a flat 
potential, we now find static models for all values of $b$ and all $\gamma\in[0,2]$.
%%%%%%%%%%%%%%%%%%%%%%%%%%%%%%%%%%%%%%%%%%%%%%%%%%%%%%%%%%%%%%%
\section{Bianchi I models with an exponential potentials} \label{sec:Bianchi}
%%%%%%%%%%%%%%%%%%%%%%%%%%%%%%%%%%%%%%%%%%%%%%%%%%%%%%%%%%%%%%%
We now turn our attention to the dynamics of Bianchi type I 
models in the brane\hs world scenario with exponential potential.
This class of models is characterized by a metric of the form
\begin{equation} 
\mbox{ds}^2 = -dt^2 + \sum_{\alpha=1}^3 A^2_\alpha(t)(dx^\alpha)^2\,. 
\label{line}
\end{equation}
Again, we follow very closely the analysis done in \cite{Campos2}, 
the only difference being that we have the additional equation (\ref{gammadot}) 
describing the dynamics of the scalar field. 

As we have discussed above, the non\hs zero contributions from the 
five\hs dimensional Weyl tensor are ${\cal U}$ and ${\cal P}_{ab}$ but
since the second one has no evolution equation we will assume for 
simplicity that ${\cal P}_{ab}\sigma^{ab}=0$.  

When introducing appropriate variables, we have to consider two different 
cases: (A) ${\cal U}\geq 0$; (B) ${\cal U}\leq 0$.
%%%%%%%%%%%%%%%%%%%%%%%%%%%%%%%%%%%%%%%%%%%%%%%%
\subsection{${\cal U} \geq 0 $}
%%%%%%%%%%%%%%%%%%%%%%%%%%%%%%%%%%%%%%%%%%%%%%%%
This case is formally very similar to subsection \ref{sec:FLRWA}. 
We obtain a compact state space using the dimensionless variables 
$\gamma,\Omega_\rho,\Omega_{\Lambda},\Omega_\sigma,\Omega_\lambda,
\Omega_{\cal U}$, where
\begin{equation} 
\Omega_\sigma \equiv \frac{\sigma^2}{3H^2} \,. \label{defsigma}
\end{equation}
defines a normalized variable for the shear contribution, and the remaining 
variables are defined by (\ref{def1}) and (\ref{def2}).
The Friedmann constraint (\ref{Friedmann}) reduces to
\be
\label{friedas}
\Omega_{\rho}+\Omega_{\Lambda}+\Omega_{\sigma}+\Omega_{\lambda}+\Omega_{\cal{U}}=1\;.
\ee
Using the time derivative
\be
'\equiv |H|^{-1}d/dt 
\nonumber
\ee
the system of dynamical equations is given by
\bea
\gamma '&=&\eps \sqrt{3\gamma} (\gamma -2)[\sqrt{3\gamma}
+\eps~ b \sqrt{1-\Omega_{\sigma}-\Omega_{\Lambda}-\Omega_{\lambda}-
\Omega_{\cal{U}}}]\;,
\nonumber\\
\Omega_{\Lambda}'&=&\eps[3\gamma(1+\Omega_{\lambda}-\Omega_{\Lambda})+
(6-3\gamma)\Omega_{\sigma}+
(4-3\gamma)\Omega_{\cal{U}}]\Omega_{\Lambda}\;,
\nonumber\\
\Omega_{\sigma}'&=&\eps[(3\gamma-6)(1-\Omega_{\sigma})
+3\gamma(\Omega_{\lambda}-\Omega_{\Lambda})+(4-3\gamma)\Omega_{\cal{U}}]
\Omega_{\sigma}\;,
\nonumber\\
\Omega_{\lambda}'&=&\eps[3\gamma(\Omega_{\lambda}-\Omega_{\Lambda}-1)+
(6-3\gamma)\Omega_{\sigma}+
(4-3\gamma)\Omega_{\cal{U}}]\Omega_{\lambda}\;,
\nonumber\\
\Omega_{\cal{U}}'&=&\eps[(3\gamma-4)(1-\Omega_{\cal{U}})+(6-3\gamma)
\Omega_{\sigma}+3\gamma(\Omega_{\lambda}-\Omega_{\Lambda})]\Omega_{\cal{U}}\;.
\nonumber\\
\label{systemas}
\eea
The equilibrium points of this system, their coordinates in state space and 
their eigenvalues are given in TABLE VI.
%%%%%%%%%%%%%%%%%%%%%%%%%%%%%%%%%%%%%%%%%%%%%%
%TABLE; EIGENVALUES As
%%%%%%%%%%%%%%%%%%%%%%%%%%%%%%%%%%%%%%%%%%%%%%
\begin{table*}
\caption{\label{tab:table7} This table gives the coordinates and 
eigenvalues of the critical points with ${\cal U} \geq 0 $. 
We have defined $\zeta=\sqrt{\frac{64}{b^2}-15}$.}
\begin{ruledtabular}
\begin{tabular}{ccc}
Model  & Coordinates   & Eigenvalues \\ \hline
$\mbox{K}_\epsilon^0$ & $(0,0,1,0,0)$ & non-hyperbolic \\
$\mbox{K}_\epsilon^2(\Omega_{\sigma})$ & $(2,0,\Omega_{\sigma},0,0)$ & $
(6\epsilon+\sqrt{6}b\sqrt{1-\Omega_{\sigma}},6\eps,0,-6\eps,2\eps)$ \\
$\mbox{dS}_\epsilon^2$ & $(2,1,0,0,0)$ & $
(6\eps,-6\eps,-6\eps,-12\eps,-4\eps)$ \\
$\mbox{m}_\epsilon^0(\Omega_{\Lambda},\Omega_{\lambda})~for~ b=0$ 
\footnote{The eigenvalues of these points can only be evaluated for 
$\Omega_{\Lambda}+\Omega_{\lambda}\neq 1$. For 
$\Omega_{\Lambda}+\Omega_{\lambda}= 1$ a 
perturbative analysis has to be carried out.}& 
$(0,\Omega_{\Lambda},0,\Omega_{\lambda},0)$ & $-2\epsilon(3,0,3,0,2)$ \\
$\mbox{m}_\epsilon^0(\Omega_{\Lambda},\Omega_{\lambda})~for~ b 
\neq 0~^a$ & $(0,\Omega_{\Lambda},0,\Omega_{\lambda},0)$ & $(\infty\footnote{This actually reads 
$lim_{\gamma \rightarrow 0}(-6\eps-\sqrt{3}b\sqrt{\frac{1-\Omega_{\Lambda}-\Omega_{\lambda}}
{\gamma}})=\infty.$},0,-6\epsilon,0,-4\eps)$ \\
$\mbox{m}_\epsilon^2$ & $(2,0,0,1,0)$ & $2\epsilon(3,6,3,3,4)$ \\
$\mbox{R}_\epsilon^0$ & $(0,0,0,0,1)$ & non-hyperbolic \\
$\mbox{R}_\epsilon^2$ & $(2,0,0,0,1)$ & $2\epsilon(3,2,-1,-4,-1)$ \\
$\mbox{F}_{+}^{b^2/3}~~\footnote{$b^2\in[0,6]$}$ & $(\frac{b^2}{3},0,0,0,0)$ & 
$(\frac{b^2}{2}-3,b^2,b^2-6,-b^2,b^2-4)$\\
$\mbox{A}_{+}^{4/3}(b)~~\footnote{$b^2\in [4,\infty[$} $ & $(\frac{4}{3},0,0,0,
1-\frac{4}{b^2})$ 
& $(-\frac{1}{2}(1+\zeta),4,-2,-4,-\frac{1}{2}(1-\zeta))$\\
\end{tabular}
\end{ruledtabular}
\end{table*}
The only new equilibrium points that we obtain in this subsection are the 
anisotropic models $K$. These have been identified in \cite{Campos2} as the 
general relativistic vacuum Kasner models with line element (\ref{line}) where   
\begin{equation} 
A_\alpha = t^{2p_\alpha}~~\mbox{and}~~  \sum_{\alpha=1}^3 p_\alpha = 
\sum_{\alpha=1}^3 p^2_\alpha = 1\,. \label{kasner}
\end{equation}
%%%%%%%%%%%%%%%%%%%%%%%%%%%%%%%%%%%%%%%%%%%%%%%%
\subsection{${\cal U} \leq 0$}
%%%%%%%%%%%%%%%%%%%%%%%%%%%%%%%%%%%%%%%%%%%%%%%%
In analogy to subsection~\ref{sec:FLRWC}, we consider the variables 
$\gamma,Z,\bar{\Omega}_\rho,\bar{\Omega}_{\Lambda},\bar{\Omega}_\sigma,
\bar{\Omega}_\lambda$, where 
$\bar{\Omega}_\sigma$ is defined as $\Omega_\sigma$ in equation 
(\ref{defsigma}),
but normalized with respect to $N$ defined in (\ref{defz}) and the 
remaining variables are as in subsection~\ref{sec:FLRWC}.  With these 
variables the Friedmann constraint (\ref{Friedmann}) reads
\be
\label{friedbs}
\bar{\Omega}_{\rho}+\bar{\Omega}_{\Lambda}+\bar{\Omega}_{\sigma}+
\bar{\Omega}_{\lambda}=1\;.
\ee
With respect to the time derivative $'\equiv N^{-1}d/dt$, the dynamical
system becomes 
\begin{eqnarray}
\gamma '&=& \sqrt{3 \gamma} (\gamma -2)[\sqrt{3 \gamma}Z + b 
\sqrt{1-\bar{\Omega}_{\lambda}-\bar{\Omega}_{\Lambda}
-\bar{\Omega}_{\sigma}}]\;,
\nonumber\\
Z'&=&-[( \frac{3}{2} \gamma-2)+(3-\frac{3}{2} \gamma)
\bar{\Omega}_{\sigma}+\frac{3}{2} \gamma (\bar{\Omega}_{\lambda}-
\bar{\Omega}_{\Lambda})](1-Z^2)\;,
\nonumber\\
\bar{\Omega}_{\Lambda}'&=&[(6-3 \gamma)\bar{\Omega}_{\sigma}+3 
\gamma (1+\bar{\Omega}_{\lambda}-\bar{\Omega}_{\Lambda})] Z \bar{\Omega}_
{\Lambda}\;,
\nonumber\\
\bar{\Omega}_{\sigma}'&=&[(3 \gamma -6)(1-\bar{\Omega}_{\sigma})+
3\gamma (\bar{\Omega}_{\lambda}-\bar{\Omega}_{\Lambda})]Z 
\bar{\Omega}_{\sigma}\;,
\nonumber\\
\bar{\Omega}_{\lambda}'&=&[(6-3 \gamma)\bar{\Omega}_{\sigma}+3 
\gamma (\bar{\Omega}_{\lambda}-\bar{\Omega}_{\Lambda}-1)] 
Z \bar{\Omega}_{\lambda}\,.
\label{systembs}
\end{eqnarray}
The equilibrium points of this system, their coordinates in state space and 
their eigenvalues are given in TABLE VII.
%%%%%%%%%%%%%%%%%%%%%%%%%%%%%%%%%%%%%%%%%%%%%%%%%%%%%
%TABLE; EIGENVALUES Bs
%%%%%%%%%%%%%%%%%%%%%%%%%%%%%%%%%%%%%%%%%%%%%%%%%%%%%
\begin{table*}
\caption{\label{tab:table8} This table gives the coordinates and 
eigenvalues of the critical points with ${\cal U} \leq 0$. 
We have defined $\xi=\sqrt{64-15b^2},\phi=\sqrt{2}b(\frac{3}{2}-\frac{1}
{\gamma})\sqrt{3\gamma(1-
\bar{\Omega}_\Lambda-\bar{\Omega}_{\sigma})+3\bar{\Omega}_\sigma-2},
~\varphi=\sqrt{9\gamma^2
(\bar{\Omega}_\Lambda+\bar{\Omega}_{\sigma}-1)-18\gamma\bar{\Omega}_\sigma+
12\gamma+4}$. Notice that 
$\varphi$ is a real positive 
quantity within the allowed range of variables.}
\begin{ruledtabular}
\begin{tabular}{ccc}
Model  & Coordinates   & Eigenvalues \\ \hline
$\mbox{K}_\epsilon^0$ & $(0,\eps,0,1,0)$ & non-hyperbolic \\
$\mbox{K}_\epsilon^2(\bar{\Omega}_{\sigma})$ & $(2,\eps,0,
\bar{\Omega}_{\sigma},0)$ & 
$(6\epsilon+\sqrt{6}b\sqrt{1-\bar{\Omega}_{\sigma}},2\eps,6\eps,0,-6\eps)$ \\
$\mbox{dS}_\epsilon^2$ & $(2,\eps,1,0,0)$ & 
$(6\epsilon,-4\eps,-6\eps,-6\eps,-12\eps)$ \\
$\mbox{m}_\epsilon^0(\bar{\Omega}_{\Lambda},\bar{\Omega}_{\lambda})~
for~ b=0$ \footnote
{The eigenvalues of these points can only be evaluated for 
$\bar{\Omega}_{\Lambda}+\bar{\Omega}_{\lambda}\neq 1$. For 
$\bar{\Omega}_{\Lambda}+\bar{\Omega}_{\lambda}= 1$ a 
perturbative analysis has to be carried out.} & 
$(0,\eps,\bar{\Omega}_{\Lambda},0,\bar{\Omega}_{\lambda})$ & 
$-2\epsilon(3,2,0,3,0)$ \\
$\mbox{m}_\epsilon^0(\bar{\Omega}_{\Lambda},\bar{\Omega}_{\lambda})~
for~ b 
\neq 0~^a$ & $(0,\eps,\bar{\Omega}_{\Lambda},0,\bar{\Omega}_{\lambda})$ & 
$(\infty\footnote{This 
actually reads $lim_{\gamma \rightarrow 0}(-6\eps-\sqrt{3}b\sqrt{\frac{1-\bar
{\Omega}_{\Lambda}-\bar
{\Omega}_{\lambda}}{\gamma}})=\infty.$},-4\epsilon,0,-6\eps,0)$ \\
$\mbox{m}_\epsilon^2$ & $(2,\eps,0,0,1)$ & $2\epsilon(3,4,6,3,3)$ \\
$\mbox{nE} $ & $(\gamma^\ast,0,\bar{\Omega}^\ast_{\Lambda},
\bar{\Omega}^\ast_{\sigma},
\bar{\Omega}^\ast_{\lambda})$ & $(\phi,\varphi,0,-\varphi,0)$ \\
$\mbox{F}_{+}^{b^2/3}~~\footnote{$b^2 \in [0,6]$}$ & $(\frac{b^2}{3},1,0,0,0)$
 & $(\frac{b^2}{2}-3,b^2-4,b^2,b^2-6,-b^2)$\\
$\mbox{A}_{+}^{4/3}(b)~~\footnote{$b^2 \in [0,4]$}$ & $(\frac{4}{3},-\frac{b}
{2},0,0,0)$ & $(\frac{1}{4}(b-\xi),\frac{1}{4}(b+\xi),-2b, b , 2b)$\\
\end{tabular}
\end{ruledtabular}
\end{table*}
In addition to the anisotropic Kasner models we also obtain the set of saddle 
points $nE$ whose coordinates $(\gamma^\ast,0,\bar{\Omega}^\ast_{\Lambda},
\bar{\Omega}^\ast_{\sigma},
\bar{\Omega}^\ast_{\lambda})$  satisfy
\begin{equation}
2-\frac{3}{2}\gamma^\ast(1+\bar{\Omega}^\ast_{\lambda}-\bar{\Omega}^\ast_\sigma
-\bar{\Omega}^\ast_\Lambda)-3\bar{\Omega}^\ast_\sigma =0
\label{nEcon1}
\end{equation}
and 
\begin{equation}
\sqrt{\gamma^\ast}(\gamma^\ast -2)b
\sqrt{1-\bar{\Omega}^\ast_{\Lambda}-\bar{\Omega}^\ast_{\lambda}-
\bar{\Omega}^\ast_{\sigma}}]\,.
\label{nEcon2}
\end{equation}
Furthermore the constants ${\cal U}^\ast,{\sigma^\ast}^2$ must satisfy 
the condition
\begin{equation}
-{\cal U}^\ast \geq \frac{\lambda\kappa^2}{6}({\sigma^\ast}^2+\Lambda)
\end{equation}
in order for the energy density $\rho^\ast$ to be non\hs negative.

We can see from (\ref{nEcon1}) and (\ref{friedbs}) that there are vacuum 
solutions with $\bar{\Omega}^\ast_{\sigma}=2/3$ for all values of $\gamma^\ast$.
Furthermore we find non\hs vacuum models for $\gamma^\ast=0$ and $\gamma^\ast=2$. 
The set of static models $nE$ forms a 2\hs parameter set of equilibrium points which 
for $b=0$ degenerates into a 3\hs dimensional surface containing static 
equilibrium points for all values of $\gamma^\ast$. 

Note that the $\bar{\Omega}_{\Lambda}=0$\hs subset only contains static models 
for $\gamma^\ast\in[0,\frac{2}{3}]$ if $b\neq0$ and for $\gamma^\ast
\in[0,\frac{4}{3}]$ if $b=0$. 

For each fixed value of $\gamma^\ast$ the models $nE$ are the same as the 
ones found in \cite{Campos2} with scale factors
\[ A_\alpha(t) = \mbox{e}^{q_\alpha t}\,, \]
where $q_\alpha$ are constants which satisfy
\[ \sum_{\alpha =1}^3 q_\alpha = 0 ~~\mbox{and}~~   
   \sum_{\alpha =1}^3 q^2_\alpha = 2{\sigma^\ast}^2\,.\]
We repeat that although $H=0$, the scale factors $A_\alpha(t)$ 
are in general not static. It is only the overall scale factor 
$\prod_{\alpha=1}^3
A_\alpha(t)=1$ which remains constant. The matter can still expand or 
contract along the principal shear axis, hence forming a pancake 
singularity if one $q_\alpha$ is negative and a cigar type singularity if 
two of them are negative.  

Note that the Jacobian of the dynamical system (\ref{systembs}) is in 
general not well defined for $\gamma^\ast=0$~\footnote{For $\gamma=b=0$ 
the Jacobian is well defined and the non\hs zero eigenvalues of $nE$ are given by 
$\pm \varphi\mid_{\gamma=0}$.}. In that case a perturbative study confirms 
that the points $nE\mid_{\gamma^\ast=0}$ are also saddle points in state space.
%%%%%%%%%%%%%%%%%%%%%%%%%%%%%%%%%%%%%
\subsection{Qualitative Analysis}
%%%%%%%%%%%%%%%%%%%%%%%%%%%%%%%%%%%%%
We summarize the dynamical character of the equilibrium points obtained in 
this section in TABLE VIII below.
%%%%%%%%%%%%%%%%%%%%%%%%%%%%%%%%%%%%%%%%%%%%%%%%%%%%%
%TABLE; DYNAMICAL CHARACTER  Bs for Lambda=0
%%%%%%%%%%%%%%%%%%%%%%%%%%%%%%%%%%%%%%%%%%%%%%%%%%%%%
\begin{table*}
\caption{\label{tab:table9} Dynamical character of the critical points 
in the Bianchi I case for $\Lambda=0$.}
\begin{ruledtabular}
\begin{tabular}{ccccccc}
Model  & $b=0$ & $0<b^2<4$ & $b^2=4$ & $4<b^2<6$ & $b^2=6$ & $b^2>6$ \\ \hline
$\mbox{K}_+^0$ & saddle & source &  source & source & source & source \\
$\mbox{K}_-^0$ & saddle & saddle & saddle & saddle & saddle & saddle \\ 
$\mbox{K}_\epsilon^2(\Omega_{\sigma})$ & saddle & saddle & saddle & saddle & 
saddle & saddle \\ 
$\mbox{m}_{+}^0(\Omega_{\lambda})$ & sink & saddle & saddle & saddle & saddle 
& saddle \\ 
$\mbox{m}_{-}^0(\Omega_{\lambda})$ & source & source & source & source 
& source & source \\
$\mbox{m}_{+}^2$ & source & source & source & source & source & source \\
$\mbox{m}_{-}^2$ & sink & sink & sink & sink & sink & sink \\
$\mbox{R}_\epsilon^0$ & saddle & saddle & saddle & saddle & saddle & saddle \\ 
$\mbox{R}_\epsilon^2$ & saddle & saddle & saddle & saddle & saddle & saddle \\ 
$\mbox{nE}$ & saddle & saddle & saddle & saddle & saddle & saddle \\ 
$\mbox{F}_{+}^{b^2/3}$ & sink & sink & saddle 
\footnote{Notice that this point is an attractor for all general relativistic 
open or flat 
models, while it is a repeller for all general relativistic closed models.}  
& saddle & saddle & -\\
$\mbox{A}_{+}^{4/3}(b)$\footnote{Notice that this model has negative non-local 
energy density $\cal{U}$ for $b^2<4$, ${\cal{U}}=0$ for $b^2=4$ and 
${\cal{U}}>0$ for $b^2>4$} & saddle & saddle & saddle $^a$ & sink & sink & 
sink \\
\end{tabular}
\end{ruledtabular}
\end{table*}
%%%%%%%%%%%%%%%%%%%%%%%%%%%%%%%%%%%%%%%%%%%%%%%%%%%%%
%TABLE; DYNAMICAL CHARACTER  Bs
%%%%%%%%%%%%%%%%%%%%%%%%%%%%%%%%%%%%%%%%%%%%%%%%%%%%%
\begin{table*}
\caption{\label{tab:table10} 
Dynamical character of the critical points 
in the Bianchi I case for general $\Lambda$.}
\begin{ruledtabular}
\begin{tabular}{ccccccc}
Model  & $b=0$ & $0<b^2<4$ & $b^2=4$ & $4<b^2<6$ & $b^2=6$ & $b^2>6$ \\ \hline
$\mbox{K}_+^0$ & saddle & source &  source & source & source & source \\
$\mbox{K}_-^0$ & saddle & saddle & saddle & saddle & saddle & saddle \\ 
$\mbox{K}_\epsilon^2(\Omega_{\sigma})$ & saddle & 
saddle &  saddle & saddle & saddle & saddle \\ 
$\mbox{dS}_\epsilon^2$ & saddle & saddle & saddle & saddle & saddle & saddle\\ 
$\mbox{m}_{+}^0(\Omega_{\Lambda},\Omega_{\lambda})$ & sink & saddle & saddle 
& saddle & saddle & saddle \\ 
$\mbox{m}_{-}^0(\Omega_{\Lambda},\Omega_{\lambda})$ & source &  source & 
source  & source & source & source\\
$\mbox{m}_{+}^2$ & source & source & source & source & source & source \\
$\mbox{m}_{-}^2$ & sink & sink & sink & sink & sink & sink \\
$\mbox{R}_\epsilon^0$ & saddle & saddle & saddle & saddle & saddle & saddle \\ 
$\mbox{R}_\epsilon^2$ & saddle & saddle & saddle & saddle & saddle & saddle \\ 
$\mbox{nE}$ & saddle & saddle & saddle & saddle & saddle & saddle \\ 
$\mbox{F}_{+}^{b^2/3}$ & sink & saddle & saddle & saddle & saddle & -\\
$\mbox{A}_{+}^{4/3}(b)$\footnote{Notice that this model has negative non-local 
energy density $\cal{U}$ for $b^2<4$, ${\cal{U}}=0$ for $b^2=4$ and 
${\cal{U}}>0$ for $b^2>4$} & saddle & saddle & saddle & saddle & saddle & 
saddle \\
\end{tabular}
\end{ruledtabular}
\end{table*}
The main question we want to address here is whether the initial singularity 
is isotropic or not when restricting our analysis to
Bianchi I models. We obtain the following result. 

For $b=0$ we find that the only equilibrium points of the dynamical system 
(\ref{systemas}, \ref{systembs}) that are sources and therefore present stable 
initial configurations are the expanding non\hs general relativistic BDL model 
with $\gamma=2$ ($m_+^2$) and the line of the flat and collapsing models 
with $\gamma=0$ ($m_{-}^0(\Omega_{\Lambda},\Omega(\lambda))$) including the flat FLRW model and the 
maximally non-general relativistic BDL model. All these models are isotropic.

If $b\leq0$, the dynamical system possesses in addition to the sources above 
one  further source denoted by  $K_+^0$. This equilibrium point represents 
the anisotropic Kasner model with $\gamma=0$. 

This means that if $b=0$, the initial singularity must be isotropic, since 
the anisotropic models are not stable at early times. If on the other hand the 
potential $V(\phi)$ is not flat, i.e. $b\neq0$, we find that the initial 
singularity could be anisotropic, since the anisotropic Kasner model is stable 
at early times.
%%%%%%%%%%%%%%%%%%%%%%%%%%%%%%%%%%%%%%%%%%%%%%%%%%%%%%%%%%%%%%%%%%%%%
\section{Discussion and Conclusion} \label{sec:Conclusion}
%%%%%%%%%%%%%%%%%%%%%%%%%%%%%%%%%%%%%%%%%%%%%%%%%%%%%%%%%%%%%%%%%%%%%
We end this paper with a comparison of our results with work 
previously done in this area. In particular, we focus on two issues:  
Firstly we discuss the occurrence and stability of equilibrium points 
in the dynamical systems analysis of FLRW and Bianchi I models with and 
without an exponential potential and compare our work to that done by 
Campos and Sopuerta \cite{Campos2}. Secondly we discuss the issue of 
isotropization in the past; here we mainly comment on the results recently 
obtained by Coley {\it et al} in (\cite{coley1,coley2}).

Addressing the first issue let us state the differences between this work and 
\cite{Campos1,Campos2} where the scalar field free analog of the scenario was studied. 
Formally, the new features of our work is contained in an additional evolution 
equation for $\gamma$ (\ref{gammadot}) thus enlarging the dynamical system analyzed 
with respect to the one studied in \cite{Campos2}. This means in particular that we should 
expect to obtain no more than the equilibrium points already obtained in \cite{Campos2}. 
The stability of these equilibrium points on the other hand is expected to be altered since the 
additional dynamical equation yields an additional eigenvalue in the dynamical 
systems analysis which can potentially destabilize the equilibrium points.
In fact we found that we in general only recover the equilibrium points that 
Campos et al found for any linear barotropic equation of state for the 
special values of $\gamma=0,2$ which correspond to $\dot{\phi}<<V$ and $V<<\dot{\phi}$ 
respectively.

It is worthwhile mentioning that the "new" equilibrium points $F_+^{b^2/3}$, 
$X_+^{2/3}(b)$ and $A_+^{4/3}(b)$ which are moving in state space as the steepness 
of the potential (characterized by the value of $b$) is increased are in fact not new. 
The first point simply corresponds to the flat FLRW model with equation of state parameter 
$\gamma$ depending on the value of $b$. More interestingly, the last two correspond to the 
bifurcations occurring for $\gamma=\frac{2}{3},\frac{4}{3}$ in \cite{Campos2}: 
for each value of $b$ the model $X_+^{2/3}(b)$ corresponds to a point of the 
line of general\hs relativistic, non-static equilibrium points which occurs in 
\cite{Campos1} for $\gamma=\frac{2}{3}$ (see figure 4 in that paper), and 
similarly the model $A_+^{4/3}(b)$ corresponds for each value of $b$ to one of 
the points on the line of equilibrium points joining the models denoted by 
$F_+$ and $R_+$ for $\gamma=\frac{4}{3}$ in \cite{Campos2} (see figure 2 in that 
paper). There the stability of the equilibrium points forming the bifurcations 
was not discussed in much detail. We find that for sufficiently steep potential
 ($b^2>4$) the model $A_+^{4/3}(b)$ is a stable sink in the 
$\Lambda=0$-subset of the state space of the Bianchi I models and the 
$\Lambda=k=0$-subset of the state space of the FLRW models. This means that 
in that case this point which has scale factor $a(t)=t^{1/2}$ and 
non-vanishing positive non-local energy density $\cal U$ is a future attractor
 with the equation of state of pure radiation. It is the only non\hs collapsing 
future attractor, but it is unstable for $\Lambda,k\neq0$. 

The second important issue we want to address here is the issue of 
isotropization. In \cite{coley1} it was claimed that the initial singularity 
in the brane-world scenario is necessarily isotropic. It was shown that for 
$\gamma\geq1$ the non-general relativistic BDL-model $m_+$ \footnote{This 
model is referred to as the BRW solution ${\cal F}_b$ in 
\cite {coley1,coley2}} with scale factor $a(t)=t^{1/{3\gamma}}$ is a source 
in the state space of all Bianchi IX models, and it was also claimed that 
this is the {\em only} source with physically relevant values of 
$\gamma(\gamma\geq1)$. 
In this paper we have considered the more general situation of 
$\gamma\in[0,2]$. We have found {\em all} equilibrium points of the state space
and identified the past attractors without imposing any constraints on $\gamma$.
We confirm that the BDL model with $\gamma=2$ is a generic past attractor 
in both the FLRW case and the Bianchi I case. We want to point out however, that
in the analysis of the Bianchi I models the anisotropic Kasner model with 
$\gamma=0~(K_+^0)$ is another past attractor unless the potential is flat 
($b=0$). This model has the equation of state $p=-\rho$ and corresponds to a slow
rolling scalar field ($\dot{\phi}<<V$). In particular no energy conditions are violated 
\footnote{We also find that the line of the isotropic flat and collapsing 
models with $\gamma=0$ ($m_{-}^0(\Omega_{\Lambda},\Omega_{\lambda})$) including 
the flat FLRW model and the maximally non\hs general relativistic BDL model is a 
sink for all values of $b$ in both the FLRW case and the Bianchi I case, hence all 
of the models contained in that line are stable in the past and possible as initial 
configurations if we allow for $\gamma<1$.}. In summary, we find that 
if we adopt the assumption on the equation of state in the early universe 
$\gamma\geq1$, the initial singularity must be isotropic. The benefit of our 
approach is that we first find {\em all} configurations that are stable in the 
past in a transparent and complete analysis. We can then exclude certain ones 
on physical grounds.

We conclude with the following remark on the future attractors in this 
scenario. In \cite{Paper1} we explained that the general relativistic models 
$F_+^{b^2/3}$ and $X_+^{2/3}(b)$ correspond to the equilibrium points obtained
in \cite{Burd} and \cite{Halliwell} in the general relativistic analysis, and
found that the stability of these models is not altered in the brane-world 
extension with vanishing non-local energy density $\cal U$. In this paper we 
find that these models are also stable against perturbations in $\cal U$. 
It is the perturbations in $\Lambda$ that grow and hence destabilize these 
models. This confirms the intuitive idea that the cosmological 
constant dominates at late times since we showed that allowing for non\hs zero 
$\Lambda$ destabilizes all expanding future attractors with $\Omega_\Lambda=0$. 
It does not destabilize the contracting non\hs general relativistic future attractor 
$m_-^2$, since at high energies the $\Lambda$ term is negligible with respect to the 
$\rho^2$ term corresponding to $\Omega_\lambda$. Hence the re\hs collapsing BDL model 
with stiff matter ($\gamma=2$) is the unique future attractor for the models discussed
in this paper.     

%%%%%%%%%%%%%%%%%%%%%%%%%%%%%%%%%%%%%%

%%%%%%%%%%%%%%%%%%%%%%%%%%%%%%%%%%%%%%

\end{document}